\documentclass[journal,10pt]{IEEEtran}

\usepackage{graphicx}
\usepackage{epstopdf}
\usepackage{amsthm,amsmath,amssymb}
\usepackage{mathrsfs}
\usepackage{setspace}
\usepackage{autobreak}
\usepackage{amsmath}
\usepackage{booktabs}
\usepackage{enumerate}
\usepackage{cite}
\usepackage{stfloats}
\usepackage{subfigure}

\usepackage{algorithm}
\usepackage{algorithmicx}
\usepackage{algpseudocode} 
\usepackage{multirow}
\usepackage{flafter}
\usepackage{bm}

\usepackage{url}
\usepackage{xcolor}
\usepackage{subfigure}
\usepackage{fancyhdr}
\usepackage{mdwmath}
\usepackage{mdwtab}
\usepackage{caption}
\usepackage{amsthm}
\usepackage{setspace}
\usepackage{tabularx}
\usepackage{tabu}
\usepackage{lipsum}

\newtheorem{proposition}{Proposition}
\newtheorem{remark}{Remark}

\allowdisplaybreaks
\makeatletter
\def\ScaleIfNeeded{%
\ifdim\Gin@nat@width>\linewidth \linewidth \else \Gin@nat@width
\fi } \makeatother
\begin{document}

\title{Machine Learning Empowered Resource Allocation in IRS Aided MISO-NOMA Networks}

\author{
{Xinyu~Gao,~\IEEEmembership{Student Member,~IEEE,}
Yuanwei~Liu,~\IEEEmembership{Senior Member,~IEEE,}\\
Xiao~Liu,~\IEEEmembership{Student Member,~IEEE,}
and Lingyang~Song,~\IEEEmembership{Fellow,~IEEE}

\thanks{Part of this work has been presented at the IEEE Global Communications Conference, Dec. 7-Dec. 11, 2020~\cite{IEEEhowto:XinyuGC}.}
\thanks{X. Gao, Y. Liu and X. Liu are with Queen Mary University of London, London, UK (email:\{x.gao, yuanwei.liu, x.liu,\}@qmul.ac.uk).}
}
\thanks{L. Song is with the Department of Electronics Engineering, Peking University, Beijing, China (e-mail: lingyang.song@pku.edu.cn).
}
}

\maketitle

\begin{abstract}
A novel framework of intelligent reflecting surface (IRS)-aided multiple-input single-output (MISO) non-orthogonal multiple access (NOMA) network is proposed, where a base station (BS) serves multiple clusters with unfixed number of users in each cluster. The goal is to maximize the sum-rate of all users by jointly optimizing the passive beamforming vector at the IRS, decoding order, power allocation coefficient vector and number of clusters, subject to the rate requirements of users. In order to tackle the formulated problem, a three-step approach is proposed. More particularly, a long short-term memory (LSTM) based algorithm is first adopted for predicting the mobility of users. Secondly, a K-means based Gaussian mixture model (K-GMM) algorithm is proposed for user clustering. Thirdly, a deep Q-network (DQN) based algorithm is invoked for jointly determining the phase shift matrix and power allocation policy. Simulation results are provided for demonstrating that the proposed algorithm outperforms the benchmarks, while the throughput gain of 35\% can be achieved by invoking NOMA technique instead of orthogonal multiple access (OMA).
\end{abstract}


\IEEEpeerreviewmaketitle

\section{Introduction}

With the increasing demand for large capacity in wireless networks, the conventional multiple access schemes cannot guarantee the quality of connectivity. Therefore, pursuing spectrum efficiency has become the leading focus point in wireless networks, especially in the fifth-generation (5G) and beyond networks where data volume and access volume are exploding. Although various techniques have been invoked for improving spectrum efficiency, such as large-scale multiple-input multiple-output (MIMO) \cite{IEEEhowto:Walton,IEEEhowto:CHuang} and millimeter-wave communications \cite{IEEEhowto:Rappaport}, ultra-massive and ubiquitous wireless connectivity,  it is still far from realizing enhancing spectrum and energy efficiency.
\par
Non-orthogonal multiple access (NOMA) \cite{IEEEhowto:Liu} is a novel access technique, which adopts the non-orthogonal principle to enable multiple users to share time domain, frequency domain, or code domain resources. The most attractive one is the power-domain NOMA technique \cite{IEEEhowto:Islam}, whose core idea is to superimpose the signals of two users at different powers for exploiting the spectrum more efficiently by opportunistically exploring the users' different channel conditions. Compared to the orthogonal multiple access (OMA) scheme \cite{IEEEhowto:Liu}, which can only allocate a single radio resource to one user, the NOMA scheme can amazingly increase the total throughput of wireless networks by applying superposition coding (SC) and successive interference cancelation (SIC) at the base station (BS) and receivers, respectively. 
\par
Additionally, in order to further improve spectrum efficiency and user connectivity of NOMA wireless networks, intelligent reflecting surfaces (IRS) \cite{IEEEhowto:Zhao,E2} have great ascendancy, which can intelligently reconfigure the wireless propagation environment by integrating a large number of low-cost passive reflective elements on a planar surface, thereby significantly improve the performance of wireless communication networks. The elements on the IRS can independently reflect the incident signal by controlling its amplitude and/or phase so that the received signal from BS to users can be enhanced. Compared to some communication assisting technologies such as amplify-and-forward (AF) and decode-and-forward (DF) relays, IRS-enhanced wireless networks consume less energy. Thus, sparked by the advantages of IRSs, IRS-enhanced wireless networks have been considered as one of the candidate schemes in next-generation wireless communication systems.
\par

\subsection{Prior works}
\subsubsection{Studies on IRS-aided systems}
With the aid of IRS, both spectrum efficiency and energy efficiency of wireless networks have witnessed significant improvement. The authors of \cite{IEEEhowto:Wu} jointly designed and implemented a novel IRS-aided hybrid wireless network, which showed that the IRS can be useful to achieve significant performance enhancement in typical wireless networks when comparing to the conventional networks comprising active components only. In \cite{IEEEhowto:Yu}, an jointly active beamforming and passive beamforming design algorithm was proposed for physical layer security in wireless networks. The authors of \cite{IEEEhowto:Li} considered a multiple-input single-output (MISO)-NOMA downlink communication network for minimizing the total transmit power by jointly designing the transmit precoding vectors and the reflecting coefficient vector. Finally, significant performance gain was achieved over the conventional semi-definite programming (SDP) based algorithm. An iterative algorithm was proposed in \cite{IEEEhowto:Huang} to optimize the transmit beamforming via second-order cone program (SOCP) and the reflective beamforming via the semi-definite relaxation (SDR). It showed that the performance of the IRS-aided interference channel with the proposed algorithm can significantly outperform the conventional interference channel without IRS. In \cite{IEEEhowto:Lyu}, a new hybrid wireless network comprising both active BSs and passive IRSs was studied. It proved demonstrated the effectiveness of deploying distributed IRSs in enhancing the hybrid network throughput against the conventional network without IRS, which significantly boosts the signal power. Yu \emph{et al.} \cite{IEEEhowto:XYu} developed an inner approximation (IA) algorithm, when compared to the existing designs, which cannot guarantee local optimality, the proposed algorithm showed a better performance. Sun \emph{et al.} \cite{IEEEhowto:Sun} presented an alternating optimization method to minimize the network power consumption, which alternately solves SOCP and mixed-integer quadratic programming (MIQP) problems to update the optimization variables. The authors of \cite{IEEEhowto:ZYang} investigated the problem of resource allocation for a wireless communication network with distributed reconfigurable intelligent surfaces (RIS), the simulation results that the proposed scheme achieves up to 33\% and 68\% gains in terms of the energy efficiency in both single-user and multi-user cases compared to the conventional RIS scheme and amplify-and-forward relay scheme, respectively. The adoption of a RIS for downlink multi-user communication from a multi-antenna base station was investigated in \cite{E1}. The results showed that the proposed RIS-based resource allocation methods are able to provide up to 300\% higher energy efficiency in comparison with the use of regular multi-antenna AF relaying.

\subsubsection{Studies on IRS-aided NOMA systems}
By invoking NOMA technique in IRS-aided wireless networks\cite{IEEEhowto:SZhang}, spectrum efficiency can be further enhanced when comparing to the conventional OMA schemes, such as time-division multiple access (TDMA)\cite{IEEEhowto:Jindal} and frequency-division multiple access (FDMA)\cite{IEEEhowto:Myung}. An IRS-aided NOMA system was exploited in \cite{IEEEhowto:Mu} and a novel algorithm was further developed by utilizing the sequential rank-one constraint relaxation approach to find a locally optimal rank-one solution. Fu \emph{et al.} \cite{IEEEhowto:Fu} considered jointly optimization of the transmit beamformers at the BS and the phase shift matrix at the IRS for an IRS-empowered NOMA network, which proved that performance gain was achieved. Ding \emph{et al.} \cite{IEEEhowto:Ding} proposed a novel design of IRS-assisted NOMA downlink transmission to ensure that additional cell-edge users can also be served on these beams by aligning the cell-edge users’ effective channel vectors with the predetermined spatial directions. Thus, driven by the unique characteristics of IRSs, the performance of the NOMA network can enjoy a great improvement with the complement of IRSs. In order to optimize the rate performance and ensure user fairness, the authors of \cite{IEEEhowto:Yang1} proposed a combined-channel-strength-based user ordering scheme to decouple the user-ordering design and the joint beamforming design. Afterward, an efficient algorithm was further developed to solve the formulated non-convex problem for the cases of a single-antenna BS and a multi-antenna BS, respectively, by leveraging the block coordinated decent and SDR approach. To maximize the system throughput, a three-step novel resource allocation algorithm, a low-complexity decoding order optimization algorithm, as well as a joint optimization algorithm were proposed in \cite{IEEEhowto:Zuo} to solve the problems of the channel assignment, decoding order, power allocation, and reflection coefficient design, respectively. Zhu \emph{et al.} \cite{IEEEhowto:Zhu} proposed an improved quasi-degradation condition by using IRS, which can ensure that NOMA achieves the capacity region with high probability.

\subsubsection{Studies on machine learning (ML)-aided IRS systems} 
ML \cite{IEEEhowto:Alpaydin,IEEEhowto:He,IEEEhowto:Fan} has shown great potentials to revolutionize communication systems \cite{E3}. In \cite{IEEEhowto:Song}, deep learning (DL) technique to tune the reflections of the IRS elements in real-time was developed. Results demonstrated that the DL approach yields comparable performance to the conventional approaches while significantly reducing the computational complexity. Taha \emph{et al.} \cite{IEEEhowto:Taha} adopted the DL method for learning the reflection matrices of the IRS directly from the sampled channel knowledge without any knowledge of the IRS array geometry. It proved that the application of ML in communication systems is feasible, especially in IRS-NOMA systems. Additionally, reinforcement learning (RL) was proved to be capable of tackling dynamic environments in IRS-aided wireless networks. In order to solve the joint problem of deployment and design of the IRS, a novel decay double deep Q-network (D$^{3}$QN) for the deployment and passive beamforming design of an IRS with the aid of NOMA technology was proposed in \cite{IEEEhowto:Liu2}. Taha \emph{et al.} \cite{IEEEhowto:Taha1} proposed a novel deep reinforcement learning (DRL) framework for predicting the IRS reflection matrices with minimal training overhead. The proposed online learning framework proved that it can converge to the optimal rate with perfect channel knowledge. Huang \emph{et al.} \cite{IEEEhowto:Huang1} investigated the joint design of transmit beamforming matrix at the base station and the phase shift matrix at the IRS, by leveraging recent advances in DRL, the performance and convergence rate of the proposed algorithm were improved significantly. Khan \emph{et al.} \cite{IEEEhowto:Khan} presented a DL approach for estimating and detecting symbols in signals transmitted through IRS. A novel DRL-based secure beamforming approach was proposed in \cite{IEEEhowto:Yang} to achieve the optimal beamforming policy against eavesdroppers in dynamic environments. The post-decision state (PDS) was applied to improve the learning efficiency and the prioritized experience replay (PER) schemes were utilized to enhance the secrecy performance.

\subsection{Motivations and Contributions}
\par
Although the aforementioned research contributions have laid a foundation on solving challenges in IRS-aided wireless networks and on leveraging NOMA for improving the spectrum-efficiency of networks, the dynamic environment derived from the movement of ground mobile users is ignored in the previous research contributions. Before fully reaping the advantages of IRS and NOMA techniques, how to design the phase shift matrix of the IRS and resource allocation policy based on the mobility information of users is still challenging. In contrast to the conventional MIMO-NOMA system, decoding rate conditions need to be satisfied to guarantee successful SIC in IRS-NOMA systems. Additionally, both the active beamforming and passive phase shift design affect the decoding order among users and user clustering, which makes the decoding order design, user clustering, and passive beamforming design highly coupled.
\par
ML mainly aims for automatically analyzing the law of the data and use the law to predict the unknown data. Wireless communication seeks to complete the reliable, efficient, and safe transmission of signals. In the research process, people mainly focus on the signal design at the source, the estimation of the transmission channel, and the design of receivers. As the environment continues to change, the unknown information in the communication transmission process becomes non-trivial to estimate. ML algorithms can perform deep-level feature mining on existing data and make a prediction of future changes to resist possible unknown interference. In the IRS-assisted system, IRS is used as a passive auxiliary element in signal transmission. The IRS has to update the state timely based on the variable of the source and receiver. In this case, ML is capable of overcoming the unknown information. Among many ML algorithms, reinforcement learning algorithms are capable of controlling agents that can act autonomously in a certain environment and constantly improve their behaviors through interaction with the environment. As mentioned in the IRS-assisted system, the phase adjustment of the IRS also needs to be changed accordingly by observing the objects in the environment. Thus, the RL algorithm can be invoked in the IRS-assisted system.
\par
Sparked by the above background, we aim to find the maximum sum-rate in the downlink IRS-aided MISO-NOMA network. Our contributions are summarized as follows:
\begin{itemize}
\item We propose a novel framework for IRS-NOMA aided wireless network, where an IRS is employed to enhance spectrum efficiency by proactively reflecting the incident signals.
\item We conceive users' positions and clustering based on long short-term memory (LSTM) algorithm and a K-means-based Gaussian mixture model (K-GMM), respectively. LSTM can predict the future positions for all users, while K-GMM is a general clustering algorithm to depict the distribution of users, which aims to explore optimal clustering methods under different scenarios.
\item We demonstrate that the DQN algorithm is relatively suitable for joint phase shift and resource allocation design. In contrast to the conventional Q-learning model, the DQN algorithm is capable of overcoming the memory explosion caused by a large amount of data input. Its performance can be further improved by selecting \emph{$\epsilon$} value.
\end{itemize}

\subsection{Organization}
\par
The rest of this paper is organized as follows. Section II presents the system model and problem formulation. In Section III, we propose an efficient algorithm for jointly position prediction, clustering, and resource allocation to obtain the maximum sum-rate. Section IV presents the numerical results and Section V concludes this paper.
\par
\emph{Notation:} Scalars, vectors, and matrices are denoted by lower-case, bold-face lower-case, and bold-face upper-case letters, respectively. $\mathbb{C}^{K \times N}$ denotes the space of $K \times N$ complex-valued vectors. The conjugate transpose of vector $\pmb{a}$ and inversion of matrix $\pmb{A}$ are denoted by $\pmb{a}^{H}$ and $\pmb{A}^{-1}$. diag($a$) denotes a diagonal matrix with the elements of vector $a$ on the main diagonal. $\mid\mid \pmb{a} \mid\mid$ denotes the norm of vector $\pmb{a}$. $\odot$ denotes the exclusive OR (XOR) operation. exp($\pmb{A}$) and log($\pmb{A}$) represent an exponential function with natural constant $e$ as base and a logarithmic function with a constant base of 10 for matrix $\pmb{A}$, respectively. $p(\pmb{A}|\pmb{B})$ indicates the probability of occurrence of matrix $\pmb{A}$ under the condition that another matrix $\pmb{B}$ has already occurred.

\section{System Model and Problem Formulation}

\subsection{System Model}

\begin{figure}[ht]
\centering  
\includegraphics[height=1.6in,width=3.4in]{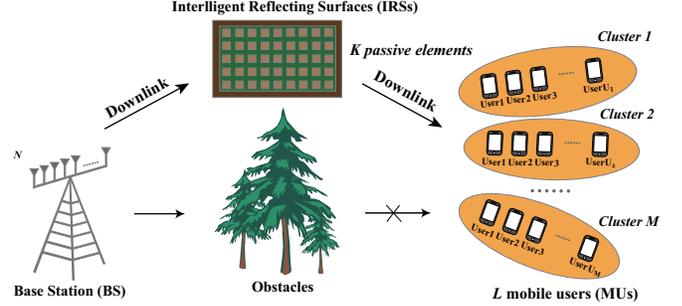}
\caption{Illustration of the MISO IRS-NOMA system.}
\label{system_model}  
\end{figure}
\par
As shown in Fig.~\ref{system_model}, we consider a downlink multi-cluster system, which consists \emph{N} transmitting antennas BS, an IRS with \emph{K} low-cost passive elements, and a fixed number \emph{L} of single antenna mobile users (MU).  In practice, the IRS is managed by BS through a smart controller, and all the users are partitioned into \emph{M} (\emph{$N \geq M$}) clusters. Addtionally, the direct transmission link between the BS and users is blocked by obstacles, and NOMA technique is employed in each cluster to improve the spectrum efficiency and quality-of-service (QoS). In the light of the system description, there are two individual channels, namely BS-IRS link and IRS-MU link, are denoted by \pmb{\emph{G}} $\in$ $\mathbb {C}$$^{K \times N}$, \emph{$\pmb{h}_{m,l_{i}}^{H}$} $\in$ $\mathbb {C}$$^{1 \times K}$, where \emph{$u_{m,l_{i}}, i=1,2,\cdots,m$} denotes the \emph{$l_{i}$}-th user in \emph{m}-th cluster. With respect to the IRS, denote \emph{$\pmb{\Theta} = {\rm diag}(\phi_{1},\phi_{2},\cdots,\phi_{k},\cdots,\phi_{K})$} as the reflection coefficients matrix of the IRS, where \emph{$\phi_{k} = \beta_{k}e^{j\theta_{k}}$}, \emph{k=1,2,3,$\cdots$,K}, \emph{$\beta_{k}$} denotes the amplitude with a constant value, such as \emph{$\beta_{k}$} = 1, while \emph{$\theta_{k}$} denotes phase of \emph{k}-th element in the IRS. Note that, in this paper, we consider both ideal case and non-ideal case of IRS \cite{IEEEhowto:Liang}. For non-ideal case, \emph{B} is denoted as the resolution bits, which can be expressed as
\begin{align}\label{1}
\mid \phi_{k} \mid^{2} = 1, \theta_{k}\in \{ \frac{2\pi n}{2^{B}},n=0,1,2,\cdots,2^{B}-1\}.
\end{align}
\par
Mark \emph{$x_{m} = \alpha_{m,l_{1}}s_{m,l_{1}} + \alpha_{m,l_{2}}s_{m,l_{2}} + \cdots + \alpha_{m,l_{m}}s_{m,l_{m}}$} as the transmit signal of the \emph{m}-th cluster, where \emph{$\{s_{m,l_{1}}, s_{m,l_{2}},\cdots, s_{m,l_{m}}\}$} and \emph{$\{\alpha_{m,l_{1}}, \alpha_{m,l_{2}},\cdots, \alpha_{m,l_{m}}\}$} denote as the transmission information and the power allocation coefficient for users \emph{$u_{m,l_{i}},i=1,2,\cdots,m$}, and \emph{$\sum_{i=1}^{m} \alpha_{m,l_{i}} = 1$}. Thus, the received signal at \emph{$u_{m,l_{i}}$} is expressed as
\begin{align}\label{2}
  y_{m,l_{i}} = \pmb{h}_{m,l_{i}}^{H} \pmb{\Theta} \pmb{G} \sum\limits_{m=1}^{M}\pmb{\omega}_{m} x_{m} + n_{m,l_{i}},
\end{align}
where \emph{n$_{m,l}$} $\sim$ $\mathcal{CN}$(0,\emph{$\delta$$^{2}$}) indicates the additive white Gaussian noise (AWGN) at user \emph{$l_{i}$} of cluster \emph{m} with zero mean and variance \emph{$\delta$$^{2}$}. Then, \emph{$\pmb{\omega}_{m}$} is represented as the corresponding beamforming vector for the \emph{m}-th cluster. The intra-cluster interference and inter-cluster interference should be considered since users are tried to employ SIC to transimit signals, and stronger users are attempted to remove the interference from the weaker users. Thus, the equation \eqref{2} can be rewritten as \eqref{3},
\begin{figure*}[!t]
\normalsize 
\begin{align}\label{3}
y_{m,l_{i}} = &\underbrace{\pmb{h}_{m,l_{i}}^{H} \pmb{\Theta} \pmb{G} \pmb{\omega}_{m} \alpha_{m,l_{i}}s_{m,l_{i}}}_{{\rm Desired\ signal}} + \underbrace{\pmb{h}_{m,l_{i}}^{H} \pmb{\Theta} \pmb{G} \pmb{\omega}_{m} \sum\limits_{j=1,j \neq i}^{m}\alpha_{m,l_{j}}s_{m,l_{j}}}_{{\rm intra-cluster\ interference}} + \underbrace{\pmb{h}_{m,l_{i}}^{H} \pmb{\Theta} \pmb{G} \sum\limits_{\gamma=1, \gamma\neq m}^{M}\pmb{\omega}_{\gamma} x_{\gamma}}_{{\rm inter-cluster\ interference}} + n_{m,l_{i}},
\end{align}
\hrulefill \vspace*{0pt}
\end{figure*}
where \emph{$\pmb{h}_{m,l_{i}}^{H} \pmb{\Theta} \pmb{G} \pmb{\omega}_{m} \sum_{j=1,j \neq i}^{m}\alpha_{m,l_{j}}s_{m,l_{j}}$} denotes the intra-cluster interference from other users in the same cluster and \emph{$\pmb{h}_{m,l_{i}}^{H} \pmb{\Theta} \pmb{G} \sum_{\gamma=1, \gamma\neq m}^{M}\pmb{\omega}_{\gamma} x_{\gamma}$} represents the inter-cluster interference from other clusters. Regarding weaker users, the interference introduced by stronger user won't be considered, which can be regarded as naught for expression. For beamforming matrix \emph{$\omega_{m}$}, perfect channel knowledge is assumed on the BS, thus, the zero-forcing (ZF)-based precoding method is assumed for pratical use, whose constraint can be expressed as
\begin{align}\label{4}
  \left\{
    \begin{array}{lr}
      \tilde{\pmb{h}}_{\gamma}^{H}\pmb{\Theta}\pmb{G}\pmb{\omega}_{m} = 0, \gamma \neq m, \forall \gamma \in \{1,2,3,\cdots,M\}, &  \\
      \tilde{\pmb{h}}_{m}^{H}\pmb{\Theta}\pmb{G}\pmb{\omega}_{m} = 1, \forall m \in \{1,2,3,\cdots,M\},
    \end{array}
  \right.
\end{align}
where \emph{$\tilde{\pmb{h}}_{m} = [\pmb{h}_{m,l_{1}},\pmb{h}_{m,l_{2}},\cdots,\pmb{h}_{m,U_{m}}]$} denotes a combination of of all user channels in the m-th cluster, and \emph{$U_{m}$} represents the number of users in \emph{$m$}-th cluster. Then, we denote \emph{$\pmb{H}^{H} = \overline{\pmb{h}}_{M}^{H}\pmb{\Theta}\pmb{G}$}, in which \emph{$\overline{\pmb{h}}_{M}^{H} = [\tilde{\pmb{h}}_{1},\tilde{\pmb{h}}_{2},\cdots,\tilde{\pmb{h}}_{m},\cdots,\tilde{\pmb{h}}_{M}]^{H}$} denotes a combination of channels for all $M$ clusters. Therefore, the transmit precoding beamforming vector is given by
\begin{align}\label{5}
  \pmb{W} = [\pmb{\omega}_{1},\pmb{\omega}_{2},\cdots,\pmb{\omega}_{m},\cdots,\pmb{\omega}_{M}] = \pmb{H}(\pmb{H}^{H}\pmb{H})^{-1}.
\end{align}
\par
Then, denote \emph{$\pmb{h}_{m,l_{i}}^{H} \pmb{\Theta} \pmb{G} = \pmb{\upsilon}^{H}\pmb{\Phi}_{m,l_{i}}$}, where \pmb{$\Phi$} = diag($\pmb{h}_{m,l_{i}}^{H}$)\pmb{G}, $\pmb{\upsilon} = [\upsilon_{1}, \upsilon_{2}, \cdots, \upsilon_{k}]^{H}$ where $\upsilon_{k} = e^{j\theta_{k}}$. Let $\Omega_{p,q}$ denotes the decoding order for user $q$ in cluster $p$. For instance, if $\Omega_{p,q} = a$, then user $q$ of cluster $p$ is the $a$-th signal to be decoded. Given two users \emph{$l_{i}$} and \emph{$l_{j}$} in the \emph{m}-th cluster, whose decoding order meets the condition \emph{$\Omega_{m,l_{i}} < \Omega_{m,l_{j}}$}, so the signal-to-interference-plus-noise ratio (SINR) of user \emph{$u_{m,l_{i}}$} to decode its own signal given by
\begin{align}\label{6}
\tau_{m,l_{i} \rightarrow m,l_{i}} = \frac{\mid\pmb{\upsilon}^{H}\pmb{\Phi}_{m,l_{i}}\pmb{\omega}_{m}\alpha_{m,l_{i}}\mid^2}{\sum\limits_{\Omega_{m,l_{\overline{j}}}>\Omega_{m,l_{i}}}\mid\pmb{\upsilon}^{H}\pmb{\Phi}_{m,l_{i}}\pmb{\omega}_{m}\alpha_{m,l_{\overline{j}}}\mid^2+\delta^{2}}.
\end{align}
and the SINR for \emph{$u_{m,l_{j}}$} to decode the received signal \emph{$s_{m,l_{i}}$} is expressed as
\begin{align}\label{7}
\tau_{m,l_{j} \rightarrow m,l_{i}} = \frac{\mid\pmb{\upsilon}^{H}\pmb{\Phi}_{m,l_{j}}\pmb{\omega}_{m}\alpha_{m,l_{i}}\mid^2}{\sum\limits_{\Omega_{m,l_{\overline{j}}}>\Omega_{m,l_{i}}}\mid\pmb{\upsilon}^{H}\pmb{\Phi}_{m,l_{j}}\pmb{\omega}_{m}\alpha_{m,l_{\overline{j}}}\mid^2+\delta^{2}}.
\end{align}
\par
It is worth pointing out that all the users have to meet QoS requirements and guarantee success SIC under a given decoding order. Therefore, for any give user \emph{$i$} and \emph{$j$} in the \emph{m}-th cluster with \emph{$\Omega_{m,l_{j}} > \Omega_{m,l_{i}}$}, the following constraint has to be satisfied
\begin{align}\label{8}
\tau_{m,l_{i} \rightarrow m,l_{i}} \geq \tau_{m,\tilde{l}_{i}},
\end{align}
where \emph{$\tau_{m,\tilde{l}_{i}}$} represents the minimum received SINR that has to achieve. Also, according to the \emph{$R={\rm log}_{2}(1+{\rm SINR})$}, it can be written as
\begin{align}\label{9}
  R_{m,l_{i} \rightarrow m,l_{i}} \geq R_{m,\tilde{l}_{i}}.
\end{align}
\par
Therefore, among all the users, for any given two users \emph{$l_{i}$} and \emph{$l_{j}$} in the \emph{m}-th cluster, \emph{$\Omega_{m,l_{i}} < \Omega_{m,l_{j}}$}, the success rate fairness conditions can be expressed as
\begin{align}\label{10}
  R_{m,l_{j} \rightarrow m,l_{i}} \geq R_{m,l_{i} \rightarrow m,l_{i}}.
\end{align}

\begin{proposition}\label{proposition 1}
  For any two users \emph{$l_{i}$} and \emph{$l_{j}$} in \emph{m}-th cluster, given the optimal decoding order \emph{$\Omega_{m,l_{j}} > \Omega_{m,l_{i}}$}, $R_{m,l_{j} \rightarrow m,l_{i}} \geq R_{m,l_{i} \rightarrow m,l_{i}}$ is the necessary condition for $R_{m,l_{i} \rightarrow m,l_{i}} \geq R_{m,\tilde{l}_{i}}$.
  \begin{proof}
    See Appendix~A.
  \end{proof}
\end{proposition}
\textbf{Proposition 1} indicates that when the optimal decoding order of NOMA is given, the constraint \eqref{9} can be removed, while the performance of NOMA networks will not be affected.

\subsection{Problem Fomulation}
In this paper, our goal is to design a novel protocol for achieving the maximum sum-rate of all users in the IRS-aided MISO-NOMA network by jointly optimizing the passive beamforming vector \emph{$\pmb{\upsilon}$} at the IRS, decoding order \emph{$\pmb{\Omega}$}, power allocation coefficient vector \emph{$\pmb{\alpha}$}, the number of clusters \emph{M}, subject to the rate requirements at \emph{L} users. Thus, the optimization problem is formulated as
\begin{align}
  \max_{\pmb{\Omega},\pmb{\upsilon},\pmb{\alpha},M} \hspace*{1em}&\sum\limits_{m=1}^{M}\sum\limits_{i=1}^{U_{m}}R_{m,L_{i}} \label{11}\\
  {\rm s.t.} \hspace*{1em}&  R_{m,l_{j} \rightarrow m,l_{i}} \geq R_{m,l_{i} \rightarrow m,l_{i}} \tag{\ref{11}{a}}, \label{11a}\\
  &\sum\limits_{m=1}^{M} \mid\mid \pmb{\omega}_{m} \mid\mid^{2} \leq \mathcal{P}, \tag{\ref{11}{b}} \label{11b}\\
  &\mid \phi_{k} \mid^{2} = 1, \theta_{k}\in [0,2\pi) \tag{\ref{11}{c}}, \label{11c}\\
  &\pmb{\Omega} \in \Pi \tag{\ref{11}{d}}, \label{11d}
\end{align}
where \emph{$U_{m}$}, \emph{$\mathcal{P}$} and \emph{$\Pi$} represents the number of users in \emph{m}-th cluster, the total transmit power, and all possible decoding orders. Constraint (\ref{11a}) guarantees that the SIC can be performed successfully, constraint (\ref{11b}) is the total transmission power constraint. Constraint (\ref{11c}) represents the considered IRS assumption. Finally, constraint (\ref{11d}) denotes the set of all the possible decoding orders. However, problem (\ref{11}) is a non-convex problem even with convex set \emph{$\pmb{\upsilon}$} since the goal of deploying and designing the IRS is for maximizing the long-term benefits, the conventional non-convex optimization algorithms that rely on statistical models may fail to promote to all the scenarios, where the varying of positions of all users can significantly differ from statistical prediction. Thus, the formulated problem falls into the field of the machine learning algorithms. In the following, we propose the machine learning-based scheme to find the solution.

\section{Proposed solutions}
\par
The conventional non-convex optimization methods tend to handle the scenario for objects with low complexity and stability, however, it cannot be a proper tool to solve problems in fast-varying environments. In this paper, we invoke LSTM and K-Gaussian mixture model (GMM) for user position estimation and clustering and then combine with the DQN algorithm to assist phase optimization of IRS.

\subsection{Positions estimation and clustering}
The resource allocation for users depends on the users' clustering. However, the users' positions change continuously with time flows, determining users' clustering becomes challenging. Here, we invoke the DL algorithm to determine the positions for users, and K-GMM for clustering. The implementation process is as follows.

\subsubsection{Positions estimation based on deep learning method}
Without loss of generality, the positions set \emph{$\pmb{L}_{t_{0}}$} of \emph{L} users are randomly generated for positions estimation by employing the acceptance-rejection sampling method proposed by Von Neumann \cite{IEEEhowto:Neumann}. Then, in order to evaluate the positions variety of users in a period of time, a recurrent neural network (RNN) which is based on the Markov hypothesis is considered for predicting data. Compared to conventional RNN algorithms, LSTM is the special representative for handling the sequence data, which can effectively avoid gradient disappearance and gradient explosion during long sequence training. Next, \emph{$\pmb{L}_{t_{0}}$} is adopted for LSTM model training, while it is also defined as initial positions for users at \emph{$t_{0}$}. Starting from the initial time \emph{$t_{0}$}, the positions \{\emph{$\pmb{L}_{t_{1}}, \pmb{L}_{t_{2}},\cdots,\pmb{L}_{t_{s}}$}\} are predicted for the corresponding time slot \{\emph{$t_{1}, t_{2},\cdots,t_{s}$}\}.
\par
According to the LSTM structure, it is obtained that the forward propagation of the algorithm has been through \emph{$z_{t}$} and \emph{$r_{t-1}$} to perform the hidden state \emph{$r_{t}$} at the next moment. \emph{$z_{t}$} is the current input and \emph{r$_{t-1}$} is the hidden state of the current moment. However, In contrast to the RNN algorithm, input gate \emph{$i_{t}$}, forget gate \emph{$f_{t}$}, output gate \emph{$o_{t}$} and internal memory unit \emph{$c_{t}$} are added inside each layer. Thus, its forward propagation formulas are
\begin{align}\label{12}
c_{t} = f_{t} \odot c_{t-1} + i_{t} \odot \tilde{c}_{t},
\end{align}
\vspace{-0.5cm}
\begin{align}\label{13}
r_{t} = o_{t} \odot {\rm Tanh}(c_{t}),
\end{align}
where \emph{$\tilde{c}_{t}$ = {\rm Tanh(}W$_{c}$x$_{c}$ + V$_{c}$r$_{t-1}$}) and \emph{$\odot$} represents XOR operation. The Tanh(\emph{$\centerdot$}) denotes the activation function, its output is between -1 and 1, which is consistent with the feature distribution centered on 0 in most scenarios. Besides, the \emph{i$_{t}$}, \emph{f$_{t}$}, \emph{o$_{t}$} are given by
\begin{align}\label{14}
i_{t} = \varphi (W_{i}x_{t} + V_{i}r_{t-1} + b_{i}),
\end{align}
\vspace{-0.5cm}
\begin{align}\label{15}
f_{t} = \varphi (W_{f}x_{t} + V_{f}r_{t-1} + b_{f}),
\end{align}
\vspace{-0.5cm}
\begin{align}\label{16}
o_{t} = \varphi (W_{o}x_{t} + V_{o}r_{t-1} + b_{o}).
\end{align}
\par
Among them, the input gate \emph{i$_{t}$} is obtained by inputting the \emph{z$_{t}$} and the hidden layer output \emph{r$_{t-1}$} of the previous step to perform a linear transformation. Then, the activation function $\varphi$ is obtained. The result of the input gate \emph{i$_{t}$} is a vector, where each element is a real number between 0 and 1. It is used to control the amount of information flowing through the valve in each dimension. \emph{$w_{i}$}, \emph{$V_{i}$} and \emph{b$_{i}$} are the parameters of the input gate, which are learned during the training process. Forget gate \emph{f$_{t}$} and output gate \emph{o$_{t}$} are calculated in the same way as input gate, with their respective parameters \emph{W}, \emph{V} and \emph{b}. \emph{$\tilde{c}_{t}$} is the state of the current candidate memory unit. Different from the RNN, the state \emph{c$_{t-1}$} of the memory unit cannot necessarily depend entirely on the state calculated by the activation function, which from the previous moment to the state \emph{c$_{t-1}$} of the current memory unit. The detailed pseudo code is shown in \textbf{Algorithm~\ref{LSTM}}.

\begin{remark}\label{remark 1}
  Parameter adjustment of LSTM network has to avoid undertraining and overfitting, otherwise, the training network fails to predict. The solution is that: 1) In the case of insufficient training, it can be achieved by adding nodes in the network or increasing the training period of the network. 2) In the case of over-fitting, it can be reduced or controlled the training cycle more, and stop the training of the network before the inflection point of the data appears.
\end{remark}

\begin{algorithm}[htbp]
  \caption{LSTM algorithm for positions predictions of users}
  \label{LSTM}
  \begin{algorithmic}[1]
  \Require LSTM network structure, generated initial positions \emph{$\pmb{L}_{t_{0}}$}, time slot evaluation flag \emph{$\overline{s}$}.
  \Ensure The positions \{\emph{$\pmb{L}_{t_{1}}, \pmb{L}_{t_{2}},\cdots,\pmb{L}_{t_{s}}$}\} at each timeslot.
  \State \textbf{Initialize:}
  Parameters of LSTM network, end time slot \emph{$t_{s}$}, \emph{L} MUs.
  \State Input set \emph{$\pmb{L}_{t_{0}}$} as training samples \emph{$\pmb{N}$} and obtained trained mature model.
  \State Define flag \emph{$\overline{s} = 0$}.
  \Repeat
  \If {Time slot flag \emph{$\overline{s}$} $<$ \emph{$s$}}
    \State Predict \emph{$\pmb{L}_{t_{\overline{s}}}$} positions for users.
    \State \emph{$\overline{s} = \overline{s} + 1$}.
    \State \emph{$\pmb{N} = \pmb{L}_{t_{0}} + \pmb{L}_{t_{\overline{s}}}$}.
    \State Input training samples \emph{$\pmb{N}$} for re-training the LSTM model.
  \EndIf
  \Until \emph{$\overline{s}$} $=$ \emph{$s$}.
  \end{algorithmic}
\end{algorithm}

\subsubsection{User clustering based on K-GMM}
In this subsection, in order to degrade the complexity of the conventional exhaustive search algorithm and extend the generality of the results, we propose a K-GMM algorithm, which combines the advantages of K-means and GMM clustering methods. The essence of the k-means model is that it draws a circle with the center of each cluster, and the maximum Euclidean distance from the midpoint of the cluster to the center of the cluster is the radius. However, the clusters (circles) fitted by the k-means model are very different from the real data distribution (possibly ellipses). In the simplest scenario, the GMM can be clustered in the same way as the K-means model. But in complex scenes, the GMM can better reflect the reality. For example, at some timeslots, if the users are clustering as circles by K-means, which may not provide optimal results for subsequent resource allocation, if K-GMM method is employed, all the possible distributions of users can be considered and compared. Most important, the suitable clustering results will be selected by K-GMM model according to the reality. However, the GMM is sensitive to the initial value, if the selected initial value is not properly, the results obtained by the model are not representative. Therefore, we first apply the K-means algorithm for pre-training to obtain initial values. Then the GMM algorithm is employed to further optimize the results.

\begin{remark}\label{remark 2}
  The result of the k-means algorithm is that each data point is assigned to one of the clusters and the GMM gives the probability that these data points are assigned to each cluster. To a certain extent, K-means is regarded as a special case of the GMM.
\end{remark}

\par
The main idea are applying K-means to roughly estimate the cluster center, then select the cluster center as the initial mean value of the GMM, and finally estimating the parameters of the Gaussian mixture model. which are elaborated as follows.
\par
\textbf{Step 1:}
All users provide their own channel state information (CSI) feedback to the BS and the BS forms a CSI set \emph{$\pmb{\mathcal{H}}$} for all users, which can be expressed as
\begin{align}\label{17}
  \pmb{\mathcal{H}} = \{\pmb{h}_{1},\pmb{h}_{2},\cdots,\pmb{h}_{L}\}.
\end{align}
\par
Besides, in order to normalize user channels, the channel vector of all users is defined as
\begin{align}\label{18}
\tilde{\pmb{h}}_{L} = \frac{\pmb{h}_{L}}{\mid\mid \pmb{h}_{L} \mid\mid}.
\end{align}
\par
\textbf{Step 2:}
In order to achieve effective clustering results, we consider two factors in the proposed clustering method: the correlation among the user channels and the gain difference in the cluster, which can be defined as 
\begin{align}\label{19}
{\rm D}_{a,b} = \big | \big | \mid\mid\tilde{\pmb{h}}_{a}\mid\mid-\mid\mid\tilde{\pmb{h}}_{b}\mid\mid \big | \big|<\rho_{1},
\end{align}
\vspace{-0.5cm}
\begin{align}\label{20}  
{\rm C_{o}(a,b)} = \dfrac{\mid\mid\tilde{\pmb{h}}_{a}\cdot\tilde{\pmb{h}}_{b}\mid\mid}{\mid\mid\tilde{\pmb{h}}_{a}\mid\mid\cdot\mid\mid\tilde{\pmb{h}}_{b}\mid\mid}>\rho_{2},
\end{align}
where \emph{$\rho_{1}$} and \emph{$\rho_{2}$} denote the pre-defined correlation thresholds while satisfying \emph{$\rho_{1},\rho_{2} \geq 0$}. We initialize all users into \emph{M} clusters and randomly chooses \emph{M} users as the centers of clusters. Then, combined K-means method, the cluster center can be calculated as
\begin{align}\label{21}
  \tilde{\pmb{C}}_{m} = \frac{1}{\mid \pmb{C}_{m}\mid} \sum\limits_{\pmb{h}_{\overline{m}} \in \pmb{\mathcal{H}}_{m}} \pmb{h}_{\overline{m}}.
\end{align}
where \emph{$\pmb{C}_{m}$}, \emph{$\pmb{h}_{\overline{m}}$} and \emph{$\pmb{\mathcal{H}}_{m}$} as the initial cluster center of \emph{m}-th cluster, the channel of \emph{$\overline{m}$} user in the \emph{m}-th cluster and the CSI set for \emph{m}-th cluster, respectively.
\par
\textbf{Step 3:}
The GMM consists of \emph{m} Gaussian distributions, each Gaussian distribution is called a "component" and these components are linearly added together for any user \emph{$l$}, which can be expressed as
\begin{align}\label{22}
  P(\pmb{\mathcal{H}}|\pmb{\kappa}) = \sum\limits_{m=1}^{M} \Psi_{m} p(\pmb{\mathcal{H}}|\kappa_{m}),
\end{align}
where \emph{$\Psi_{m} \geq 0, \sum\Psi_{m} = 1$} denotes the weights of each Gaussian distribution, \emph{$p(\pmb{\mathcal{H}}|\kappa_{m})$} is the probability density function of the \emph{m}-th Gaussian distribution while \emph{$\kappa_{m} = (\tilde{C}_{m}, \Psi_{m}^{2})$}, so the expression of probability density is
\begin{align}\label{23}
  p(\pmb{\mathcal{H}}|\kappa_{m}) = \frac{1}{\sqrt{2\pi}\sigma_{m}}{\rm exp}(-\frac{(\pmb{\mathcal{H}} - \tilde{C}_{m})^2}{2\sigma_{m}^{2}}).
\end{align}
\par
The learning process of the GMM is to estimate all the probability density function \emph{$p(\pmb{\mathcal{H}}|\kappa_{m})$} of \emph{M} Gaussian distributions. The probability of occurrence of each observation sample is expressed as a weighted probability of \emph{M} Gaussian distributions.

\begin{remark}\label{remark 3}
  Input observation data \emph{$\pmb{\mathcal{H}}$} and \emph{M} GMMs, iteratively converge to a small number \emph{$\epsilon_{0}$} and output parameter \pmb{$\kappa$} of all the GMMs. The parameters of GMMs are derived by the EM algorithm.
\end{remark}

From equation \eqref{19} to \eqref{23}, the parameters to be estimated are \emph{$\pmb{\kappa} = \{\Psi_{1}, \Psi_{2}, \cdots, \Psi_{M}; \kappa_{1}, \kappa_{2}, \cdots, \kappa_{M}\}$} and \emph{$\kappa_{M} = (\tilde{\pmb{C}}_{M}, \Psi_{M}^{2})$}. Therefore, the \emph{3M} parameters have to be estimated in this model. The maximum likelihood estimation (MLE) method is adopted to estimate \emph{$\pmb{\kappa}$}, so that the log-likelihood function \emph{$\overline{L}(\pmb{\kappa}) = {\rm log}P(\pmb{\mathcal{H}}|\pmb{\kappa})$} of the observation data \emph{$\pmb{\mathcal{H}}$} is maximized, which can be expressed as
\begin{align}\label{24}
  \overline{L}(\pmb{\kappa}) = {\rm log}P(\pmb{\mathcal{H}}|\pmb{\kappa}) = \sum\limits_{l=1}^{L}[{\rm log}(\sum\limits_{m=1}^{M}\Psi_{m} p(\pmb{h}_{l}|\kappa_{m})].
\end{align}
\par
Since the log-likelihood function \emph{$\overline{L}(\pmb{\kappa})$} contains the logarithm of the sum, it is difficult to solve, so the EM algorithm is employed as the solution of \eqref{24}. Here, the core idea of the EM algorithm is that if \emph{$\kappa_{m}$} is known, the optimal hidden variable can be inferred from the training data, namely, the E step. Conversely, if the optimal hidden variable is known, then the maximum likelihood estimation can be performed on \emph{$\kappa_{m}$}, namely, M step. The steps for the EM algorithm are elaborated as follows.
\par
According to the \emph{m}-th Gaussian distribution model \emph{$p(\pmb{\mathcal{H}}|\kappa_{m})$} determined by the weights \emph{$\Psi_{m}$}, the data set derived from \emph{$\pmb{\mathcal{H}}$} are in the same sub-model. Define \emph{$\varepsilon_{l,m}$} to evaluate the \emph{$\pmb{h}_{l}$} in the \emph{m}-th Gaussian distribution, which can be given by \eqref{25}.
\begin{figure*}[!t]
\normalsize 
\begin{align}\label{25}
  \varepsilon_{l,m} = \left\{
    \begin{array}{lr}
      1, \textbf{{\rm user}\ \emph{l}\ {\rm is\ from\ the\ model}\ \emph{m}} &  \\
      0, \textbf{{\rm otherwise}},
    \end{array}
  l = 1,2,\cdots,L, m = 1,2,\cdots,M
  \right.
\end{align}
\hrulefill \vspace*{0pt}
\end{figure*}
Thus, the complete data (\emph{$\pmb{h}_{l},\varepsilon_{l,1},\varepsilon_{l,2},\cdots,\varepsilon_{l,M}$}) can transfer \emph{$\overline{L}(\pmb{\kappa}) = {\rm log}P(\pmb{\mathcal{H}}|\pmb{\kappa})$} into \emph{$\overline{L}(\pmb{\kappa}) = {\rm log}P(\pmb{\mathcal{H}},\pmb{\varepsilon}|\pmb{\kappa})$}, which can be calculated as
\begin{align}\label{26}
  P(\pmb{\mathcal{H}},\pmb{\varepsilon}|\pmb{\kappa}) &= \prod_{l=1}^{L}P(\pmb{h}_{l},\varepsilon_{l,1},\varepsilon_{l,2},\cdots,\varepsilon_{l,M}|\pmb{\kappa}) \nonumber\\
  &=\prod_{m=1}^{M}\Psi_{m}^{\overline{\varepsilon}_{m}}\prod_{l=1}^{L}[\frac{1}{\sqrt{2\pi}\sigma_{m}}{\rm exp}(-\frac{(\pmb{h}_{l} - \tilde{C}_{m})^2}{2\sigma_{m}^{2}})]^{\varepsilon_{l,m}},
\end{align}
where \emph{$L=\sum_{m=1}^{M}\overline{\varepsilon}_{m}=\sum_{m=1}^{M}\sum_{l=1}^{L}\varepsilon_{l,m}$}. Thus, \emph{$\overline{L}(\pmb{\kappa})$} can be expressed as
\begin{align}\label{27}
  {\rm log}P(\pmb{\mathcal{H}},\pmb{\varepsilon}|\pmb{\kappa}) = &\sum_{m=1}^{M}\{\overline{\varepsilon}_{m}{\rm log}\Psi_{m}+\sum_{l=1}^{L}\varepsilon_{l,m}[{\rm log}(\frac{1}{\sqrt{2\pi}}) \notag \\ &-{\rm log}\sigma_{m}-\frac{1}{2\sigma_{m}^{2}}(\pmb{h}_{l}-\tilde{\pmb{C}}_{m})^{2}]\}.
\end{align}
\par
According to the core method of the EM algorithm, the \emph{$\overline{Q}$} function can be expressed as
\begin{align}\label{28}
  &\overline{Q}(\pmb{\kappa},\pmb{\kappa}^{(\overline{i})}) = E[{\rm log}P(\pmb{\mathcal{H}},\pmb{\varepsilon}|\pmb{\kappa})|\pmb{\mathcal{H}},\pmb{\kappa}^{(\overline{i})}] \nonumber\\
  &= \sum_{m=1}^{M}\{\sum_{l=1}^{L}(E\varepsilon_{l,m}){\rm log}\Psi_{m}+\sum_{l=1}^{L}(E\varepsilon_{l,m})[{\rm log}(\frac{1}{\sqrt{2\pi}}) \notag \\
  &-{\rm log}\sigma_{m}-\frac{1}{2\sigma_{m}^{2}}(\pmb{h}_{l}-\tilde{\pmb{C}}_{m})^{2}]\}.
\end{align}
\par
\begin{algorithm}[htbp]
  \caption{K-GMM based user clustering algorithm}
  \label{KGMM}
  \begin{algorithmic}[1]
  \Require K-means algorithm structure, GMM algorithm structure, the channels of all users $\pmb{\mathcal{H}}$.
  \Ensure The parameters of GMM \emph{$\pmb{\kappa}$} = \{$\Psi_{1}$,$\Psi_{2}$,$\cdots$,$\Psi_{M}$;$\kappa_{1}$,$\kappa_{2}$,$\cdots$,$\kappa_{M}\}$(\emph{$\kappa_{m}= \mu_{m}, \Psi_{m}^{2}$}).
  \State \textbf{Initialize:} The channels of all users $\pmb{\mathcal{H}}$, distance among all users $\pmb{D}$, correlation among all users $\pmb{C}_{o}$.
  \State Randomly choose M users and rename them as center of clusters \{$\pmb{C}_{1},\pmb{C}_{2},\cdots,\pmb{C}_{M}$\}.
  \State Calculate the distance and channel correlation users with these M users and assign all users into these clusters.
  \State Recalculate the center of each cluster using equation: $\tilde{\pmb{C}}_{m} = \frac{1}{\mid \pmb{C}_{m}\mid} \sum_{\pmb{h}_{\overline{m}} \in \pmb{\mathcal{H}}_{m}} \pmb{h}_{\overline{m}}$.
  \State Solve \eqref{24} as a single sample, and assign initial value to GMM: $\left\{ \begin{array}{lr} \mu_{m} = \tilde{\pmb{C}_{m}}, & \\ \sigma_{m}^{2} = \frac{1}{U_{m}}\sum\limits_{l=1}^{U_{m}}(\pmb{h}_{l} - C_{m})(\pmb{h}_{l} - C_{m})^{T}, \end{array} \right.$
  \Repeat
  \If {$\mid\mid \kappa_{\overline{i}+1} - \kappa_{\overline{i}} \mid\mid > \tilde{\epsilon}$}
    \State Calculate the evaluation factor: $\varepsilon_{l,m} = \frac{\Psi_{m} p(\pmb{h}_{l}|\kappa_{m})}{\sum_{m=1}^{M} \Psi_{m} p(\pmb{h}_{l}|\kappa_{m})}, \forall l \in \{1,2, \cdots, L\}.$
    \State Calculate the new iteration parameters: $\left\{ \begin{array}{lr} 
      \hat{\mu}_{m} = \frac{\sum\limits_{l=1}^{L}\hat{\varepsilon}_{l,m}\pmb{h}_{l}}{\sum\limits_{l=1}^{L}\hat{\varepsilon}_{l,m}}, \hat{\Psi}_{m} = \frac{\sum\limits_{l=1}^{L}\hat{\varepsilon}_{l,m}}{L}, \hat{\sigma}_{m}^{2} = \frac{\sum\limits_{l=1}^{L}\hat{\varepsilon}_{l,m}(\pmb{h}_{l} - \mu_{m})^{2}}{\sum\limits_{l=1}^{L}\hat{\varepsilon}_{l,m}}.\end{array} \right.$
  \EndIf
  \Until \emph{$\mid\mid \kappa_{\overline{i}+1} - \kappa_{\overline{i}} \mid\mid > \tilde{\epsilon}$}.
  \end{algorithmic}
\end{algorithm}
Denote \emph{$\overline{\varepsilon}_{l,m}=\sum_{l=1}^{L}\hat{\varepsilon}_{l,m}=\sum_{l=1}^{L}E\varepsilon_{l,m}$}, the equation \ref{28} can be re-expressed as
\begin{align}\label{29}
  &\overline{Q}(\pmb{\kappa},\pmb{\kappa}^{(\overline{i})}) = \sum_{m=1}^{M}\{\overline{\varepsilon}_{l,m}{\rm log}\Psi_{m}+\sum_{l=1}^{L}\hat{\varepsilon}_{l,m}[{\rm log}(\frac{1}{\sqrt{2\pi}}) \notag \\
  &-{\rm log}\sigma_{m}-\frac{1}{2\sigma_{m}^{2}}(\pmb{h}_{l}-\tilde{\pmb{C}}_{m})^{2}]\}.
\end{align}
\par
It can be obtained the ()\emph{i+1})-th iterated parameters, which can be expressed as
\begin{align}\label{30}
  \pmb{\kappa}^{(\overline{i}+1)} = {\rm arg} \max_{\pmb{\kappa}}\overline{Q}(\pmb{\kappa},\pmb{\kappa}^{(\overline{i})})
\end{align}
\par
Respectively make the partial derivatives of \emph{$\overline{Q}(\pmb{\kappa},\pmb{\kappa}^{(\overline{i})})$} with respect to \emph{$\tilde{\pmb{C}}_{m}$}, \emph{$\sigma_{m}$}, and \emph{$\Psi_{m}$} to be 0, it can be obtained that the parameters of \emph{$\pmb{\kappa}^{(\overline{i}+1)}$}, which can be given by
\begin{align}\label{31}
\hat{\mu}_{m} = \frac{\sum\limits_{l=1}^{L}\hat{\varepsilon}_{l,m}\pmb{h}_{l}}{\sum\limits_{l=1}^{L}\hat{\varepsilon}_{l,m}},
\end{align}
\vspace{-0.4cm}
\begin{align}\label{32}
\hat{\Psi}_{m} = \frac{\sum\limits_{l=1}^{L}\hat{\varepsilon}_{l,m}}{L},
\end{align}
\vspace{-0.5cm}
\begin{align}\label{33}
\hat{\sigma}_{m}^{2} = \frac{\sum\limits_{l=1}^{L}\hat{\varepsilon}_{l,m}(\pmb{h}_{l} - \mu_{m})^{2}}{\sum\limits_{l=1}^{L}\hat{\varepsilon}_{l,m}}.
\end{align}
\par
Repeat the calculation of EM algorithm, when \emph{$\mid\mid \kappa_{\overline{i}+1} - \kappa_{\overline{i}} \mid\mid < \tilde{\epsilon}$} is satisfied, the judgment is converged, which means that the parameters of the GMM is obtained and user clustering is finished. See \textbf{Algorithm~\ref{KGMM}} for the detail.

\subsection{Phase shift design based on deep Q-network model}
In this subsection, the deep Q-network-based algorithm for phase adjustment of IRS is discussed. According to the formulated problem, the signal blockage issues can be handled by deploying and designing the IRS. The ML method is adopted to solve the scenario where the user's positions change dynamically and pursue maximum long-term benefits. Q-learning is a value-based algorithm, which indicates that at a certain moment, taking action is expected to obtain income and the environment will feedback the corresponding reward according to the action of the agent. Thus, the core idea of the Q-network algorithm is to determine the agent, state space \emph{$\pmb{S}$}, action space \emph{$\pmb{A}$}, and reward \emph{r}. In this system model, the BS is acted as an agent, while state space and action space are denoted as \eqref{34} 
\begin{figure*}
\begin{align}\label{34}
  \left\{
    \begin{array}{lr}
      \pmb{S} = \{\theta_{1},\theta_{2},\cdots,\theta_{K};\{p_{l}\}\}, \sum\limits_{l=1}^{L}p_{l} = \mathcal{P}, &  \\
      \pmb{A} = \{[0,2\pi];\{p_{1},p_{2},\cdots,p_{v}\}\}, \{p_{v}\}\ {\rm {denotes\ available\ allocated\ power\ set}},
    \end{array}
  \right.
\end{align}
\hrulefill \vspace*{0pt}
\end{figure*}
and reward can be expressed as
\begin{align}\label{35}
  r = \sum\limits_{l=1}^{L}\sum\limits_{\overline{e}=1}^{s}(R_{l_{t_{\overline{e}}}}-R_{l_{t_{\overline{e}-1}}})
\end{align}
\par
\begin{algorithm}[htbp]
  \caption{DQN based algorithm for the IRSs}
  \label{DQN}
  \begin{algorithmic}[1]
  \Require The DQN structure, the reply memory $\mathcal{D}$, the minibatch size $\mathcal{N}$, The parameters of GMM $\pmb{\kappa}$.
  \Ensure Q-network function and decision $\mathcal{J}$.
  \State \textbf{Initialize:} Q-network weights $\pmb{\vartheta}$, target weighted Q-network $Q^{*}(S,A)$, Q-table $Q(S,A)$, state space \pmb{S}, action space \pmb{A}, reward r, the replay memory $\mathcal{D}$, time flag $\overline{s}$.
  \State The BS randomly take action $A_{t}$, phase shift $\theta_{t}$ and power allocated factor $\theta_{v}$.
  \State Input the positions for users and clustering results.
  \State Define time flag $\overline{s}=0$.
  \Repeat
    \State Choose A from S with $\epsilon$ according to $\epsilon$-greedy policy.
    \State Execute action A, observe reward r, append to S'.
    \State Accoring to the zero-forcing precoding method and calculated corresponding matrix.
    \State Determine the decoding order of clusters.
    \State Store transition (S, A; r; S') in $\mathcal{D}$.
    \State Sample random minibatch of transition $(S_{w}, A_{w}; r_{w}; S_{w}')$ from $\mathcal{D}$.
    \State Set $y_{w} = r_{W} + \beta \cdot {\rm max}_{A}Q_{S_{W},A,\vartheta}^{*}$.
    \State Perform a gradient descent step on $(y - Q_{S_{w},A_{w},\vartheta})^{2}$ according to equation \eqref{38}.
    \If {$\overline{s} < s$}
      \State update state space \pmb{S}, action space \pmb{A} and Q-table $Q(S,A)$.
      \State $\overline{s} = \overline{s} + 1$.
    \EndIf
  \Until \pmb{S} is terminal.
  \end{algorithmic}
\end{algorithm}
In the Q-learning model, the state-action value function during the process of learning for the agent can be iteratively updated, which is calculated as
\begin{align}\label{36}
  Q_{S_{t},A_{t}}^{t'} = Q_{S_{t},A_{t}}^{t} + \psi (r + \beta {\rm max}_{A_{t'}} Q_{S_{t'},A_{t'}}^{t} - Q_{S_{t},A_{t}}^{t}),
\end{align}
where \emph{$\psi$} and \emph{$\beta$} represent learning efficiency and discount parameter, respectively. Therefore, the decision policy is under the principle of choosing the maximum Q value at each time slot and at the same time, maximizing the rewards in the process of optimization.
\par
However, in some cases, as the number of data increases, the memory occupied by Q-table will increase dramatically and Q-table cannot store large amounts of data due to memory problems. Therefore, in order to reduce the memory footprint, Q-table will not be applicable. To solve it, the function approximation (FA) method is invoked to introduce a function with weights \emph{$\pmb{\vartheta}$} to approximate the Q-table. Here, we need to use some supervised learning algorithms to learn, which will turn the Q-learning model into a DQN model. \textbf{Algorithm~\ref{DQN}} shows the DQN based algorithm for controlling IRS. Thus, the new Q value is re-calculated as
\begin{align}\label{37}
  Q^{*}(S,A) = E_{S'}[r+\beta{\rm max}_{A'} Q^{*}(S^{A'},A^{'})|S,A],
\end{align}
and the loss function can be calculated as
\begin{align}\label{38}
  Loss(\pmb{\vartheta}) = \sum\limits (y - Q_{S_{t},A_{t},\vartheta})^{2},
\end{align}
where \emph{y} represents the output value calculated by the current Q-value at the next timeslot, which is given by
\begin{align}\label{39}
  y = r + \beta \cdot {\rm max}_{A_{t}}Q_{S_{t},A_{t},\vartheta}^{t}.
\end{align}

\section{Numerical results}
\begin{figure*}[ht] 
  \centering
  \includegraphics[height=3.5in,width=5in]{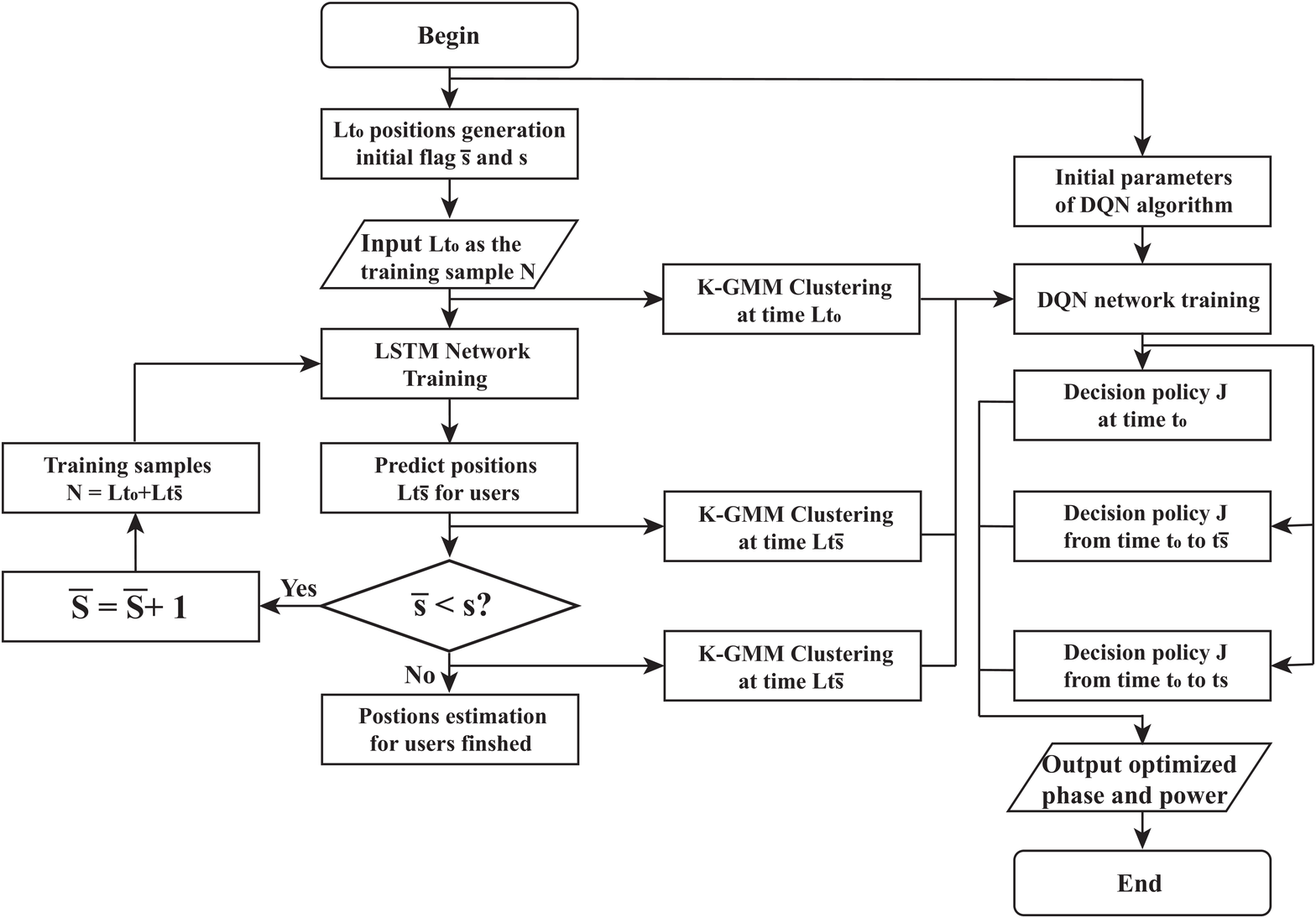}  
  \caption{System flow chart.}
  \label{System flow chart}  
  \end{figure*}
\par
In this section, numerical results are provided for verifying the effectiveness of the proposed solutions based on the system model and assumption. The flow chart of the whole process is depicted in Fig.~\ref{System flow chart}, it is obtained that the resource allocation policy depends on the requirements for the users. The path loss model is given by \emph{$\mathcal{L}(d)=\mathcal{C}d^{-\overline{\alpha}}$}, where \emph{$\mathcal{C}$} denotes the path loss when the reference distance is 1m. Additionally, the simulation parameters are shown in Tab.~\ref{Sim}, which are obtained by averaging independent channel realizations and meets the threshold requirements.
\begin{table*}[!th]
  \caption{Simulation parameters \label{Sim}} 
  \centering
  \begin{tabular}{ccc}  
  \toprule 
  Parameter & Description & Value  \\ 
  \midrule
  $\mathcal{C}$ & Path loss when d = 1m & -30dB \\
  $\delta^{2}$ & Nose power variance& -70dBW \\ 
  $\overline{\alpha}_{BU}$ & path loss factor for BS-User link& 3.5 \\ 
  $\overline{\alpha}_{IU}$ & path loss factor for IRS-User link& 2.8 \\
  $\overline{\alpha}_{BI}$ & path loss factor for BS-IRS link& 2.2 \\ 
  $\mathcal{D}$ & The replay memory capacity for DQN& 10000 \\  
  $\beta$ & Discount factor & 0.8 \\  
  $\psi$ & Learning rate & 0.1 \\  
  $\tilde{\epsilon}$ & Convergence threshold for EM algorithm & 1e-15 \\  
  $\epsilon_{0}$ & Probability decision value for $\epsilon$-greedy strategy & 0.1 \\  
  $\alpha_{0}$ & Initial power allocation coefficient for each user & 0.1 \\    
  \bottomrule
  \end{tabular}  
\end{table*}

\subsection{Positions estimation and clustering}
\begin{figure*}[htbp] 
  \centering
  \subfigure[Clustering for positions at $t_{1}$, M = 4.] {\label{4t1} \includegraphics[height=1.4in,width=2in]{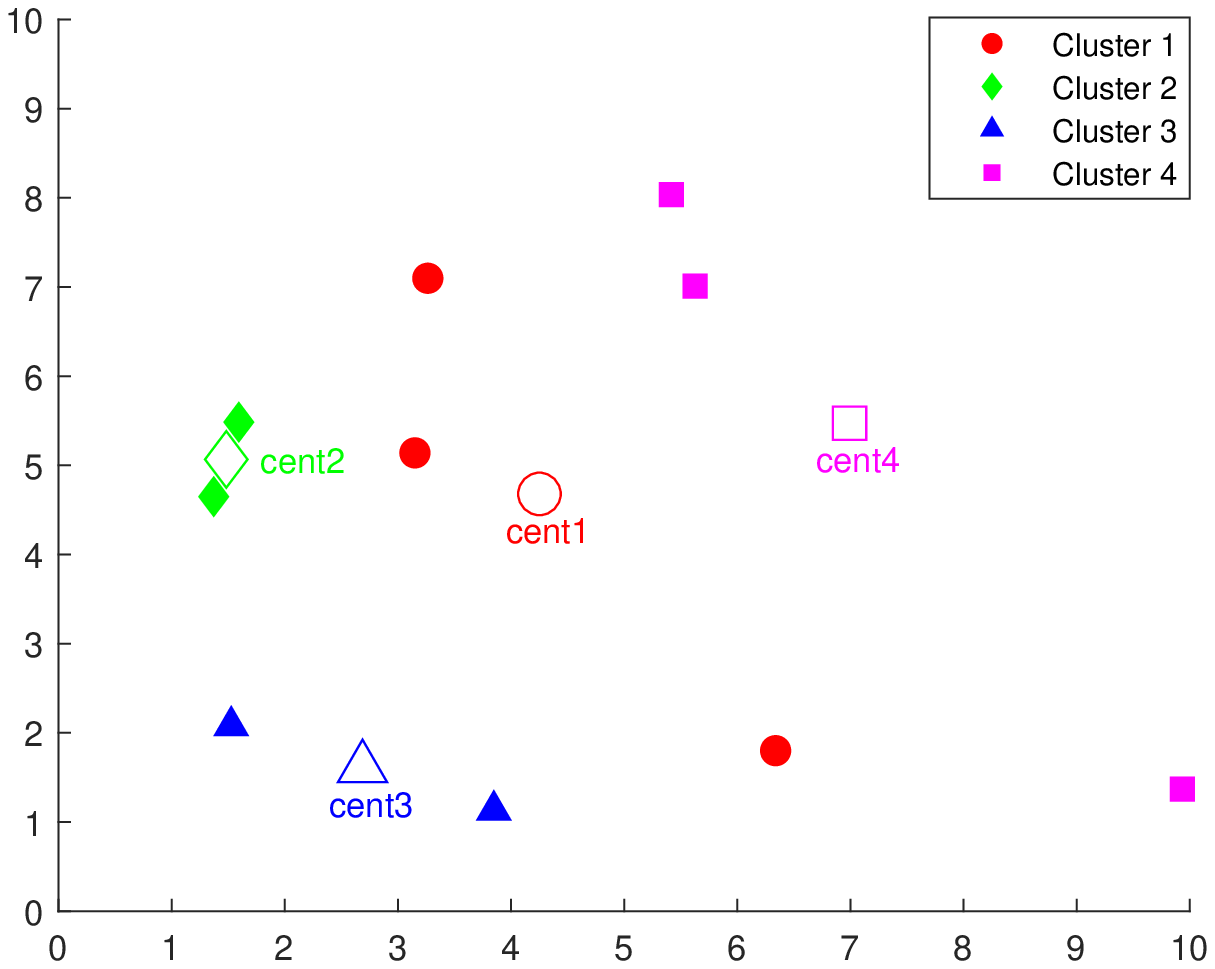}}
  \subfigure[Clustering for positions at $t_{2}$, M = 4.] {\label{4t2} \includegraphics[height=1.4in,width=2in]{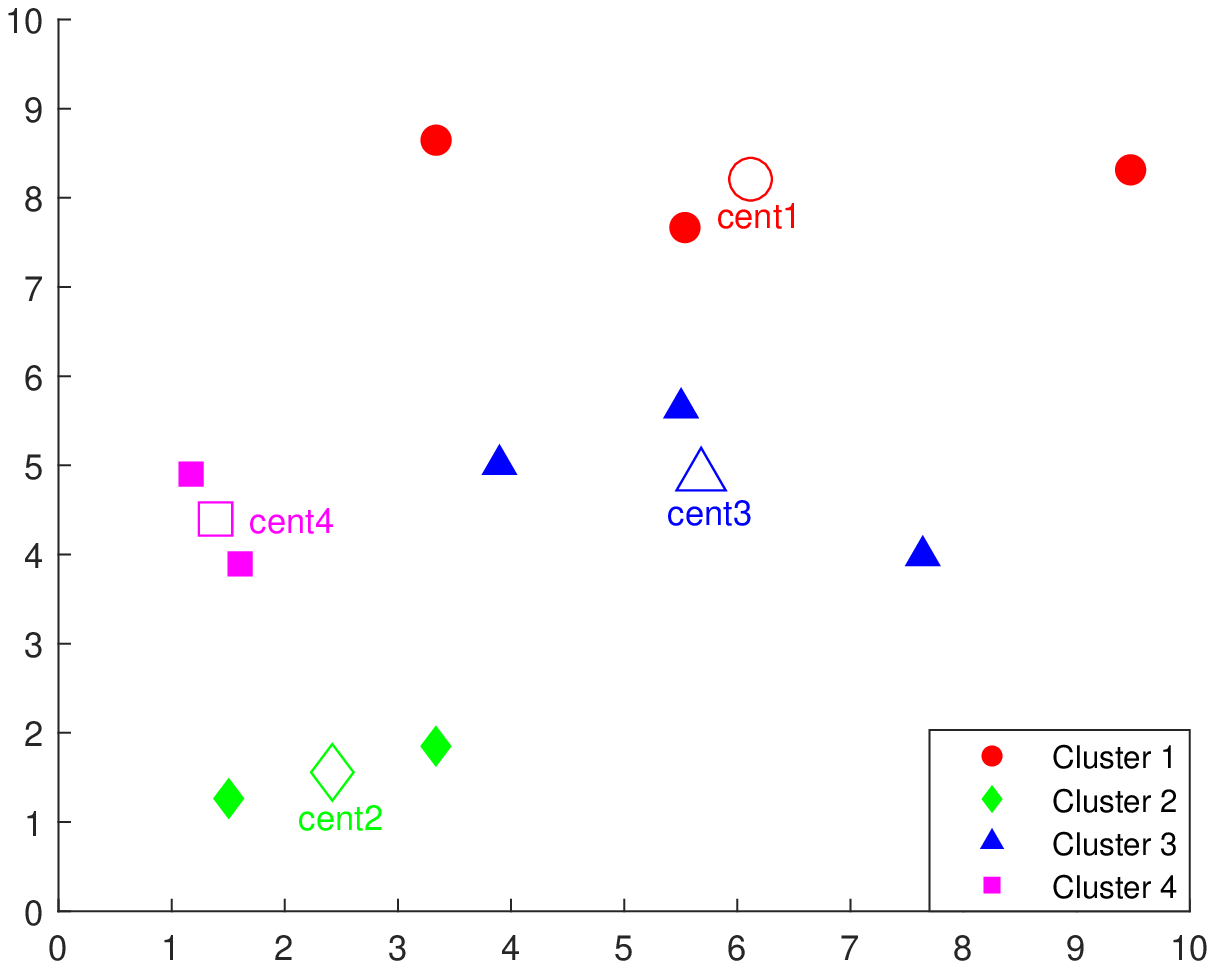}}
  \subfigure[Clustering for positions at $t_{3}$, M = 4.] {\label{4t3} \includegraphics[height=1.4in,width=2in]{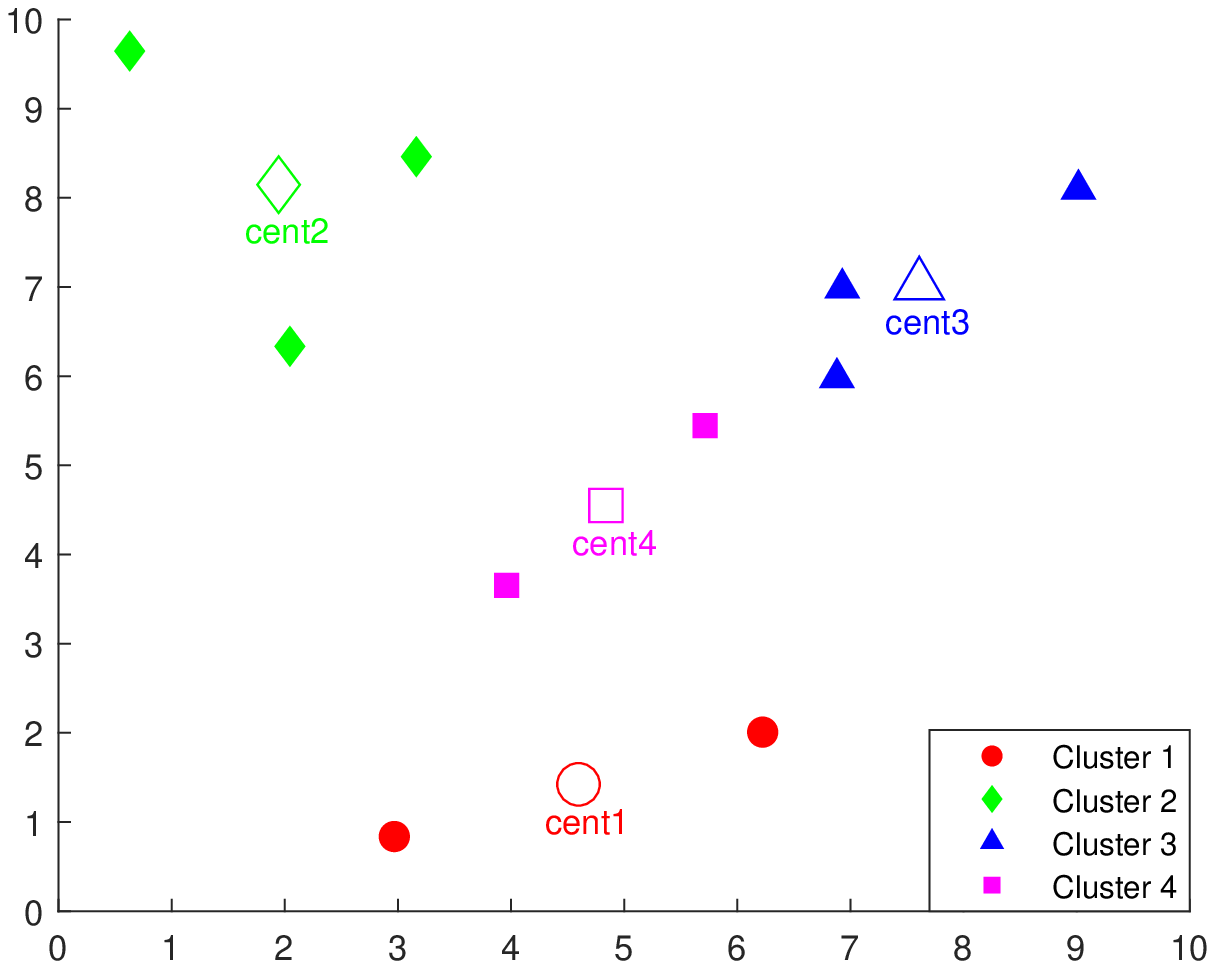}}
  \subfigure[Clustering for positions at $t_{1}$, M = 5.] {\label{5t1} \includegraphics[height=1.4in,width=2in]{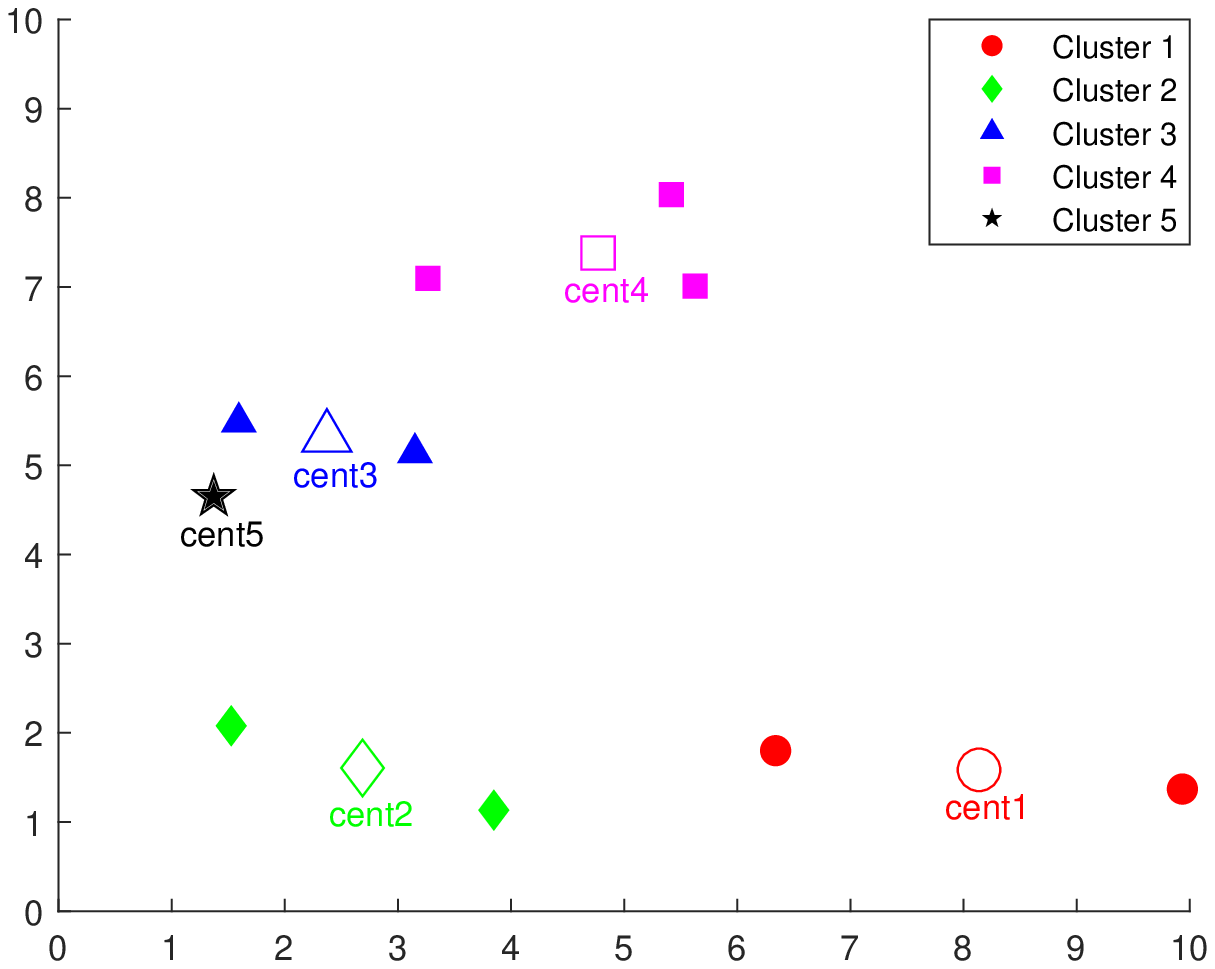}}
  \subfigure[Clustering for positions at $t_{2}$, M = 5.] {\label{5t2} \includegraphics[height=1.4in,width=2in]{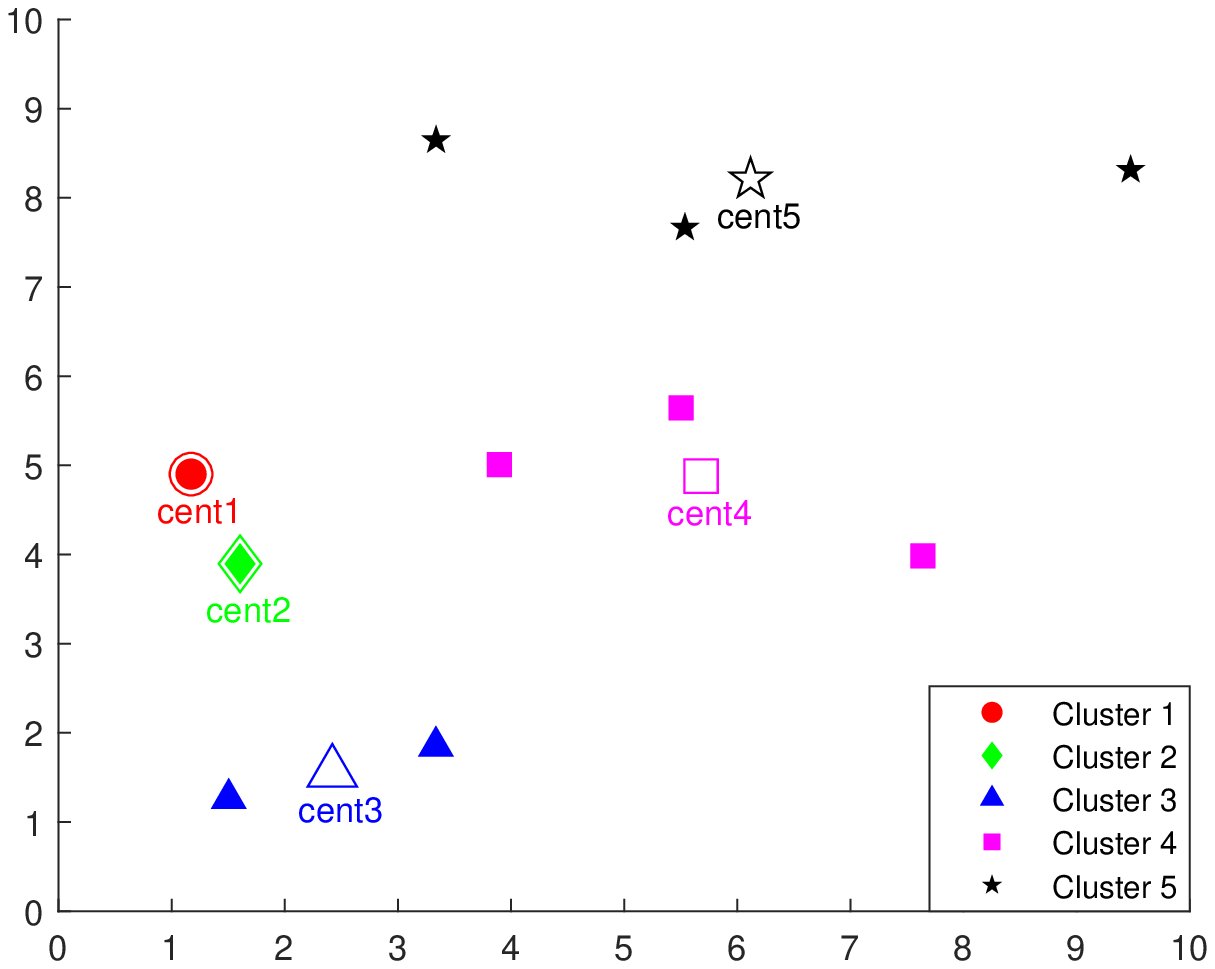}} 
  \subfigure[Clustering for positions at $t_{3}$, M = 5.] {\label{5t3} \includegraphics[height=1.4in,width=2in]{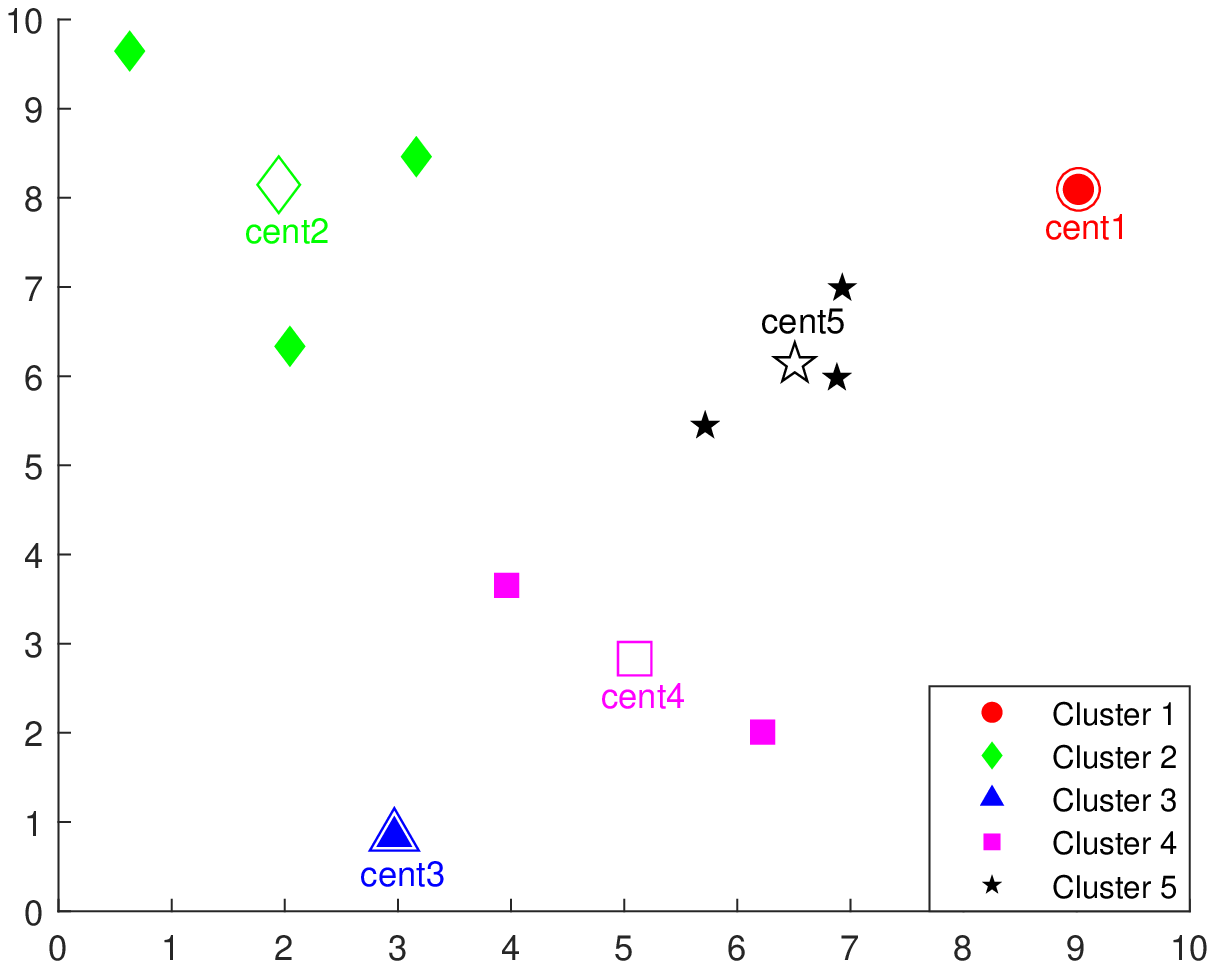}}
  \subfigure[Clustering for positions at $t_{1}$, M = 6.] {\label{6t1} \includegraphics[height=1.4in,width=2in]{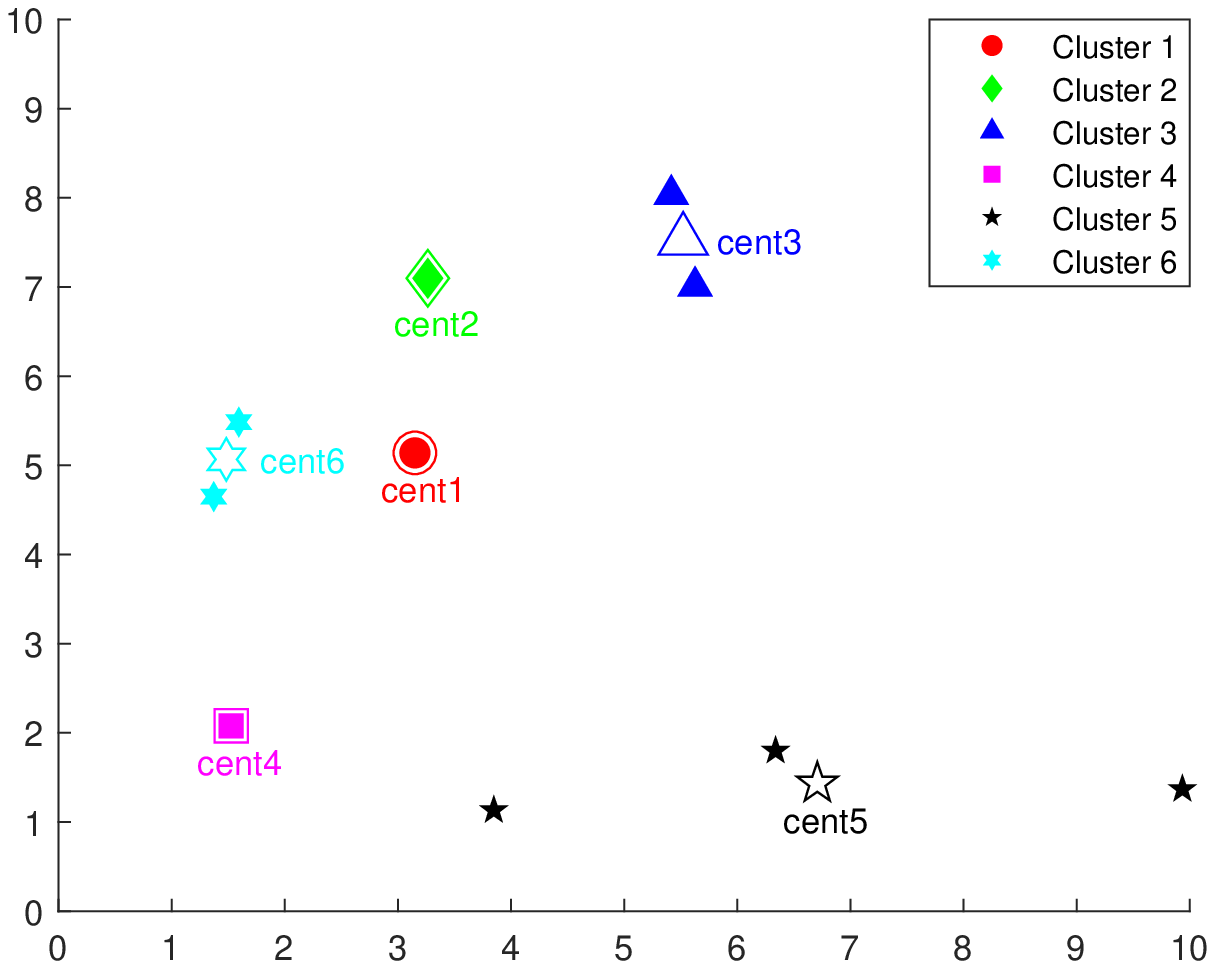}}
  \subfigure[Clustering for positions at $t_{2}$, M = 6.] {\label{6t2} \includegraphics[height=1.4in,width=2in]{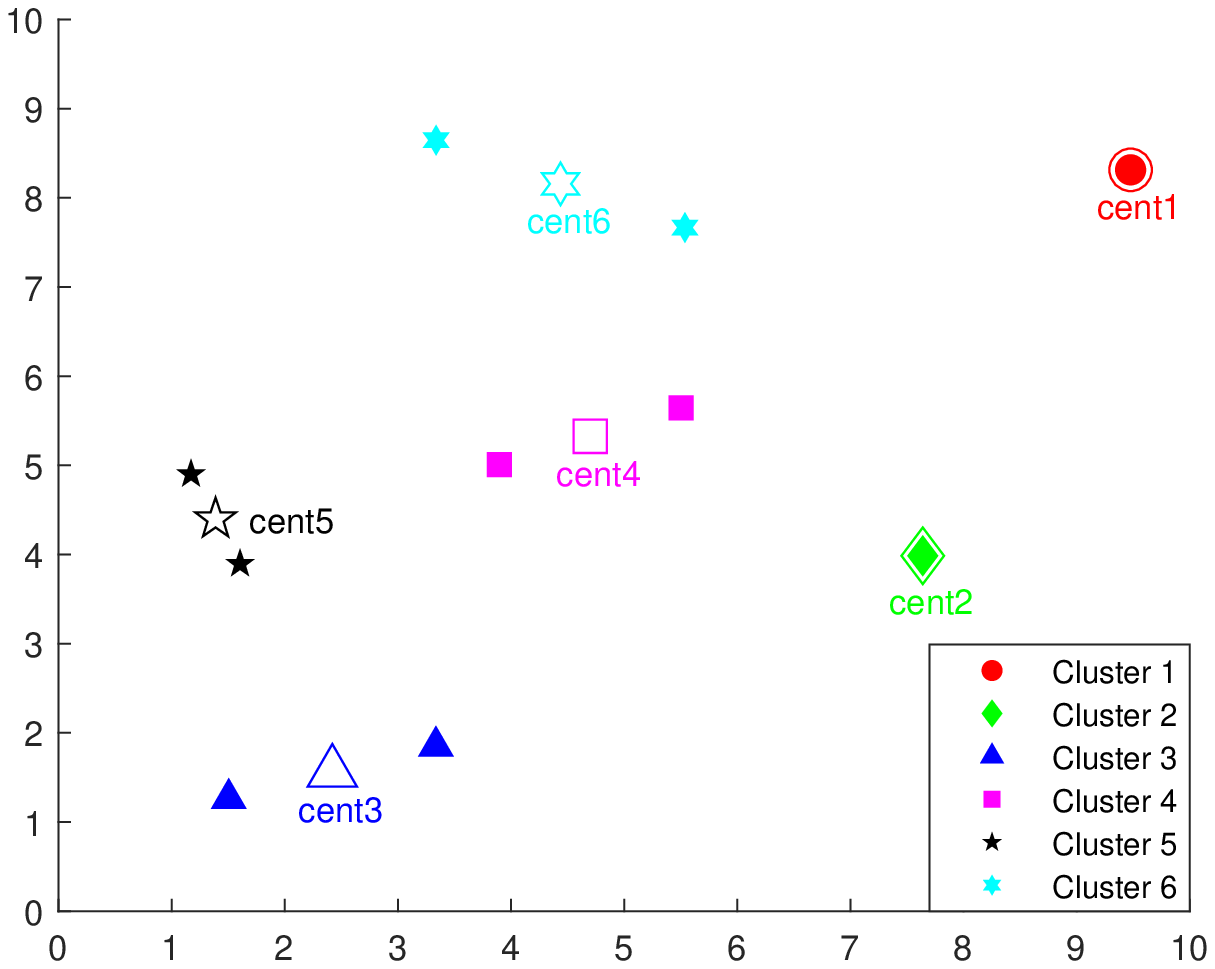}} 
  \subfigure[Clustering for positions at $t_{3}$, M = 6.] {\label{6t3} \includegraphics[height=1.4in,width=2in]{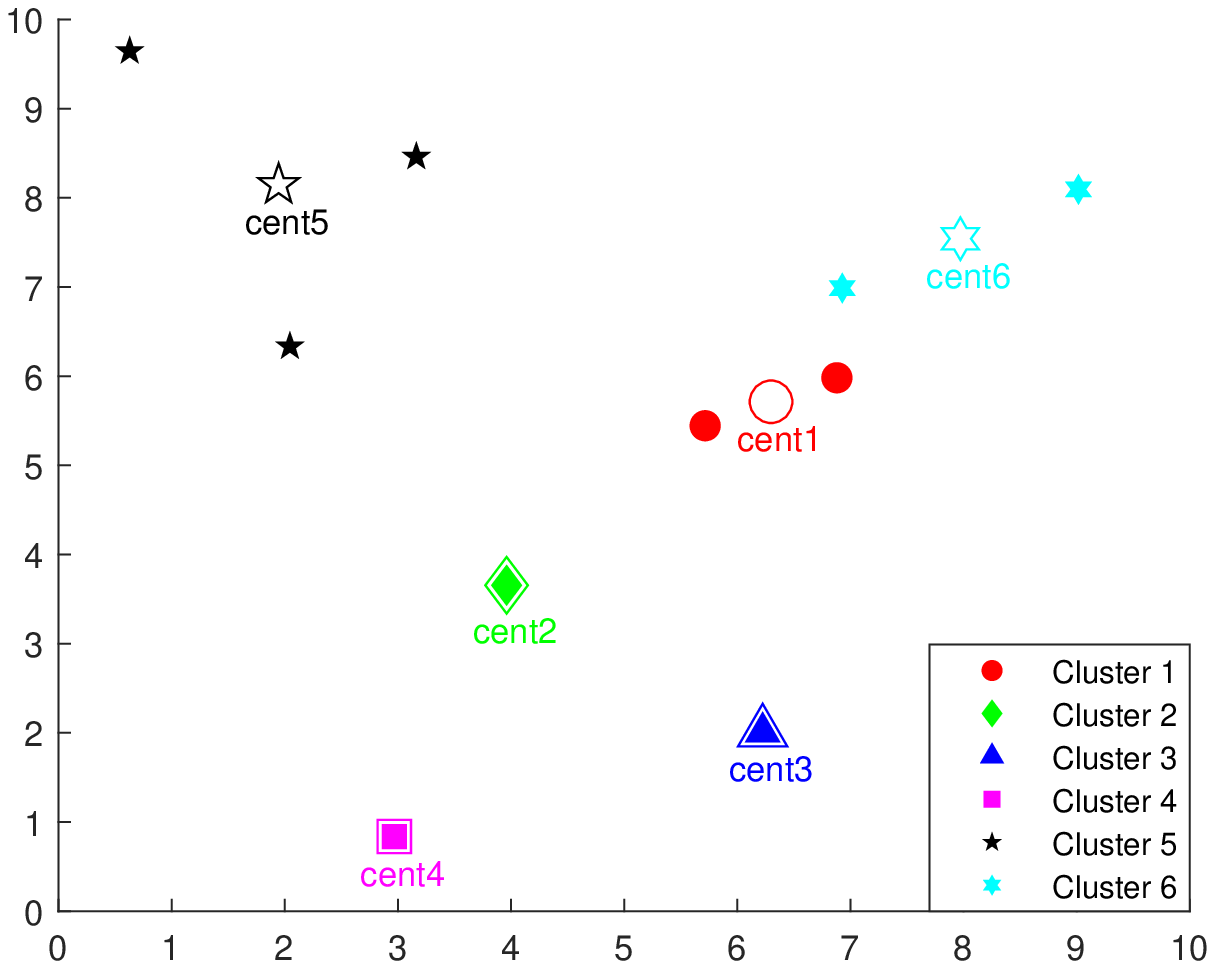}}
  \caption{Clustering for positions with different channel correlation and distance requirements among users, N=10.}
  \label{ClusteringM}       
\end{figure*}
\par
According to the illustrated system model and the definition of the solutions for position estimation and clustering, we assume that the geometric center of IRS is denoted as the origin of the coordinates to establish a Cartesian coordinate system, where the size of the IRS can be ignored compared to the distance between BS and users. For the definition of parameters for the LSTM algorithm, we define that the LSTM algorithm layer has 200 hidden units. Then, regarding training options, the solver is denoted as 'adam' and 300 training rounds are executed. To prevent gradient explosions, the gradient threshold is 1. At the same time, an initial learning rate of 0.005 is specified. After 125 training sessions, when multiplying by a factor of 0.2, the learning rate is reduced. Thus, the steps for position estimation are as follows. Firstly, employing the acceptance-rejection sampling method to generate \emph{L=10} users at time \emph{$t_{0}$}, and distribute the range of activities for all users during the whole time period. The formula for user positions generation derives \textbf{Lemma 1} in Appendix C. Then, invoking the LSTM algorithm to train the prediction model to estimate the positions for all users at time \{\emph{$t_{1},t_{2},\cdots,t_{s}$}\}. In order to ensure generality and increase the accuracy of the model prediction, the model is retrained between the adjacent timeslots, and the estimated positions at the last timeslot are included in the new training samples. Thus, the positions for users \{\emph{$\pmb{L}_{t_{1}}, \pmb{L}_{t_{2}},\cdots,\pmb{L}_{t_{s}}$}\} can be estimated. As shown in Fig.~\ref{ClusteringM}, the estimation of the positions for all users from \emph{$t_{1}$} to \emph{$t_{3}$} are depicted.
\par
The clustering results are also characterized in this figure. According to the concept of the K-GMM, the CSI of each user is assumed as known and defined as \emph{$\{\pmb{h}_{1},\pmb{h}_{2},\cdots,\pmb{h}_{L}\}$}. Since the CSI of all users obeys Rician distribution and position-dependent, the distance and correlation between any given two users can be calculated according to \eqref{19} and \eqref{20} at each timeslot \{\emph{$\pmb{L}_{t_{0}}, \pmb{L}_{t_{1}}, \pmb{L}_{t_{2}},\cdots,\pmb{L}_{t_{s}}$}\}. Thus, owing to the different predefined threshold for the distance and correlation, the \emph{L} users can be partitioned into \emph{M=3, 4, 5} clusters. In these subfigures, the different colors and shapes are adopted to represent each cluster, where the solid shape represents the user, and the hollow represents the cluster center of each cluster calculated by the EM algorithm. These centers are the positions where the passive beamforming is aligned with the launch. Additionally, in order to apply NOMA technology to the proposed system model, the number of users of each cluster is no more than three. In the subsequent section, the resource allocation policy is designed according to these three cases of clustering results.

\subsection{Performance of DQN algorithm}
Combined with the requirements, the performance of the algorithm applied is evaluated mainly from its convergence and complexity.

\subsubsection{Convergence for DQN algorithm}
\begin{figure}[ht] 
  \centering
  \includegraphics[height=2.4in,width=3.2in]{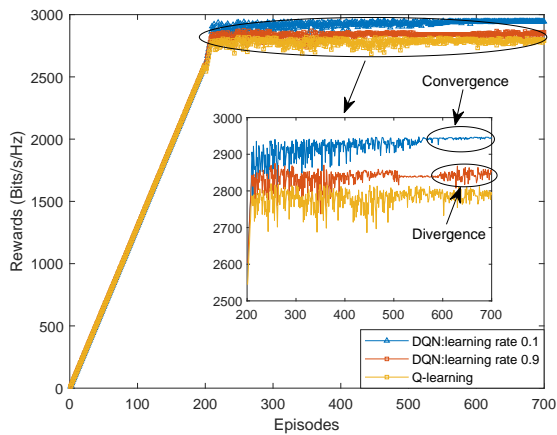}  
  \caption{Performance of DQN algorithm.}
  \label{Performance}  
  \end{figure}

From the formulated problem, our goal is to pursue the long-term rewards with dynamic positions for all users, which cannot be solved by the conventional optimization algorithm. Thus, before fully invoking the DQN algorithm, the performance of DQN should be analyzed. In Fig.~\ref{Performance}, we provide the performance for the DQN algorithm performance compared to the Q-learning algorithm, as well as the difference between the high and low learning rate, with \emph{N}=10, \emph{K}=10 and \emph{$\mathcal{P}$}=10dBm. Since the whole exploration process begins from \emph{$t_{0}$} to \emph{$t_{s}$} until achieves the optimal resource allocation policy according to positions at each timeslot, rewards are obtained based on the iterations. According to the definition of \emph{$t_{s}$}, there is an increased posture for rewards from \emph{$t_{0}$} to \emph{$t_{s}$}, at \emph{$t_{s}$}, rewards tend to a stable vibration and begin to explores whether the convergence condition is reached. In this figure, it is obtained that the Q-learning algorithm is difficult to realize the convergence under the proposed system model, while an over-defined value of learning rate conducts quicker convergence, but leads to the converged algorithm becomes divergent again.
\par
\subsubsection{Complexity for DQN algorithm}
We ignore the difference in algorithm complexity caused by different hardware and only consider the effect of the algorithm itself complexity. The complexity of the performance of the DQN algorithm also brings a huge impact on the performance of the results. As mentioned above, let \emph{$\pmb{S}$} and \emph{$\pmb{A}$} denote the state space and action space, respectively, and the definition space of policy, transition, reward model and Q function can be expressed as $\pmb{S} \times \pmb{A}$, $\pmb{S} \times \pmb{A} \times \pmb{S}$, $\pmb{S} \times \pmb{A} \times \pmb{S}$ and $\pmb{S} \times \pmb{A}$. Therefore, in the worst case, the $\pmb{S}$ and $\pmb{A}$ space definitions are too large, which will directly affect the entire DQN algorithm occupying memory, resulting in increased space complexity of the algorithm.

\subsection{Sum-rate versus Number of clusters}
\begin{figure}[ht] 
  \centering
  \includegraphics[height=2.4in,width=3.2in]{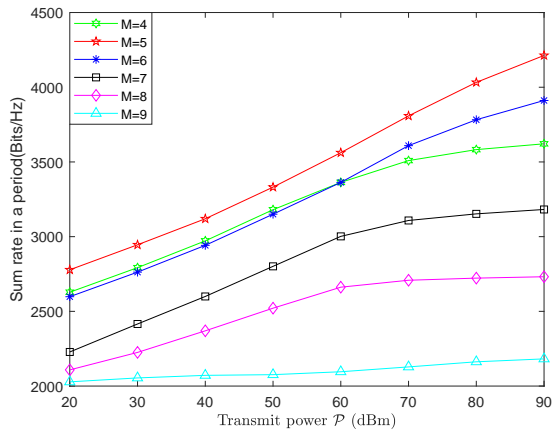}  
  \caption{The sum-rate in a period versus Number of Clusters M, K=10, N=10.}
  \label{Clusters}
  \end{figure}

In Fig.~\ref{Clusters}, we evaluate the impact of the number of clusters \emph{M} on the performance of the IRS-aided NOMA networks. The number of clusters is determined by the distance and correlation among users according to equations \eqref{19} to \eqref{20}. It can be observed that the elbow point can be found when \emph{M}=5. When the number of clusters \emph{M} from 4 to 9, the sum-rate of the proposed system monotonically upgrades with the transmit power under different numbers of clusters. This is because when the transmit power increases, the receive SINR of each user can obtain the remarkable channel gain. However, the sum-rate of each cluster is not fast improved owing to the inter-cluster interference, while the proper number of clusters can achieve the expected growth efficiency. Particularly, when the users are partitioned into 5 clusters, the sum-rate grows faster than the sum-rate for other cluster numbers. This is because when the number of clusters is too small or too large, according to the K-GMM, the difference of probability that each user belongs to each cluster will be tiny or big, which is difficult to determine the attribution of each user, then the inter-interference can be affected by the positions of clusters without only the number of clusters. However, the increasing number of clusters is also indeed to introduce more severe inter-cluster interference for the systems, especially damage the benefits of NOMA. Therefore, an appropriate number of clusters can assist reduce inter-cluster interference, as well as obtain the benefit from NOMA. Moreover, when \emph{M}=4 and \emph{M}=6, the increase rate for the sum-rate of users appears a difference while the transmit power is defined over 60dBm.
\par

\subsection{Impact of IRS}
In this subsection, we investigate the benefits brought by the IRS, there are six schemes are conceived to help demonstrate the DQN-based algorithm.
\begin{itemize}
  \item \textbf{IRS-continuous: Q-learning}: In this case, the Q-learning algorithm is selected to verify the performance of DQN, since the Q-learning is divergent compared to the DQN algorithm illustrated in Fig.~\ref{Performance}.
  \item \textbf{IRS-continuous: DQN}: In this case, the IRS phase shifters are considered as an ideal scenario, where the phase of each element can be shifted to any expected value. The DQN algorithm is included in the optimized method for phase shift, as well as the discrete scenarios for IRS phase shifters.
  \item \textbf{IRS-1bit: DQN}: In this case, the number of resolution bits of the IRS phase shifters is defined as 1, where the phase of each element can be shifted are only 0 and $\pi$.
  \item \textbf{IRS-2bit: DQN}: In this case, the number of resolution bits of the IRS phase shifters is defined as 2, where the phase of each element can be shifted are only 0, $\frac{\pi}{2}$, $\pi$ and $\frac{3\pi}{2}$.
  \item \textbf{Random phase shift}: In this case, the phase shifts of IRS elements are defined with random values.
  \item \textbf{Without IRS}: In this case, the BS serves all users without the assistant of the IRS.
\end{itemize}

\subsubsection{Sum-rate versus Total Transmit Power}
\begin{figure}[htbp]
  \centering
  \begin{minipage}[t]{0.48\textwidth}
  \centering
  \includegraphics[height=2.4in,width=3.2in]{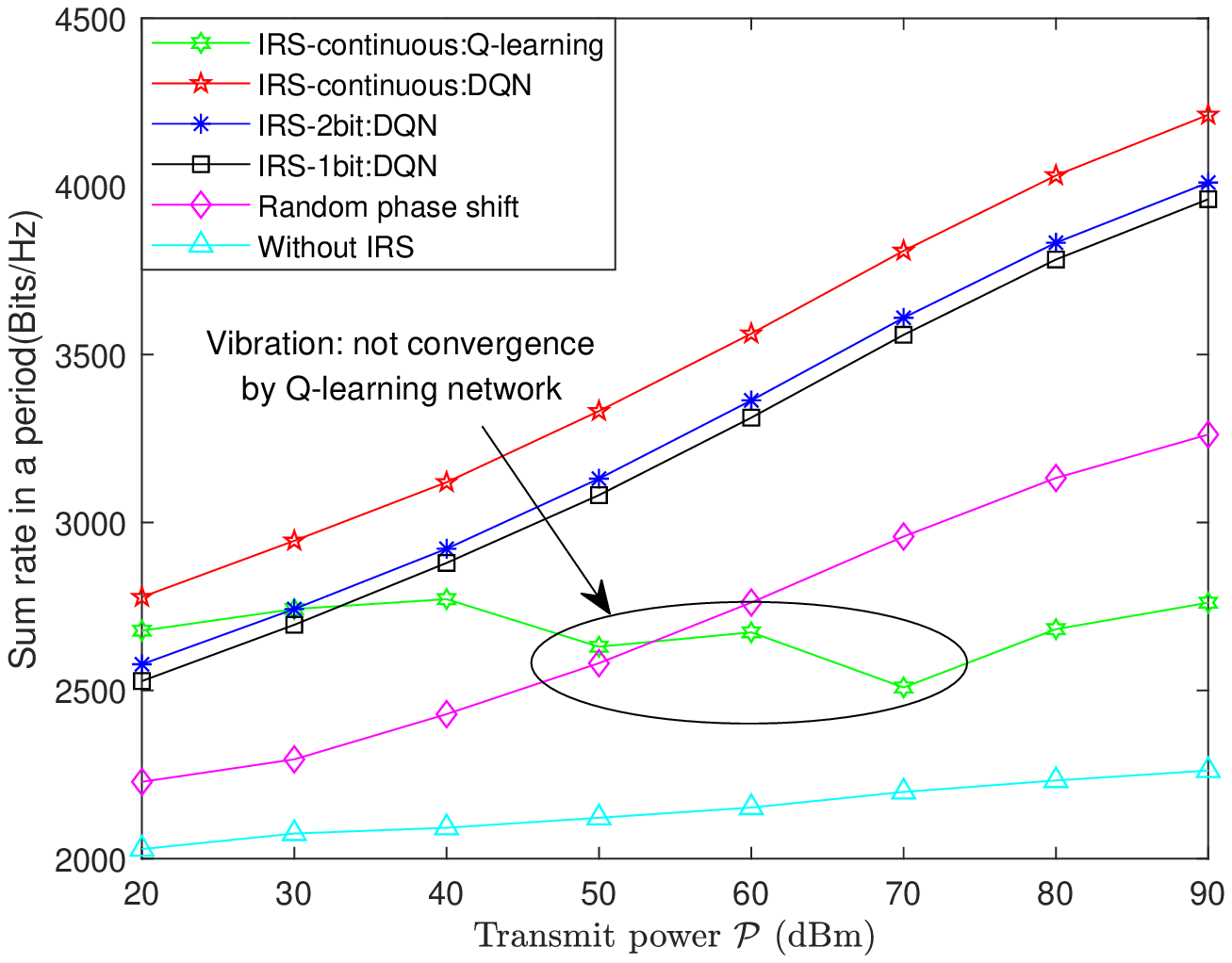}
  \caption{The sum-rate in a period versus total transmit power $\mathcal{P}$, $\emph{K}$=10, M=5, N=10.}
  \label{Power}
  \end{minipage}
  \begin{minipage}[t]{0.48\textwidth}
  \centering
  \includegraphics[height=2.4in,width=3.2in]{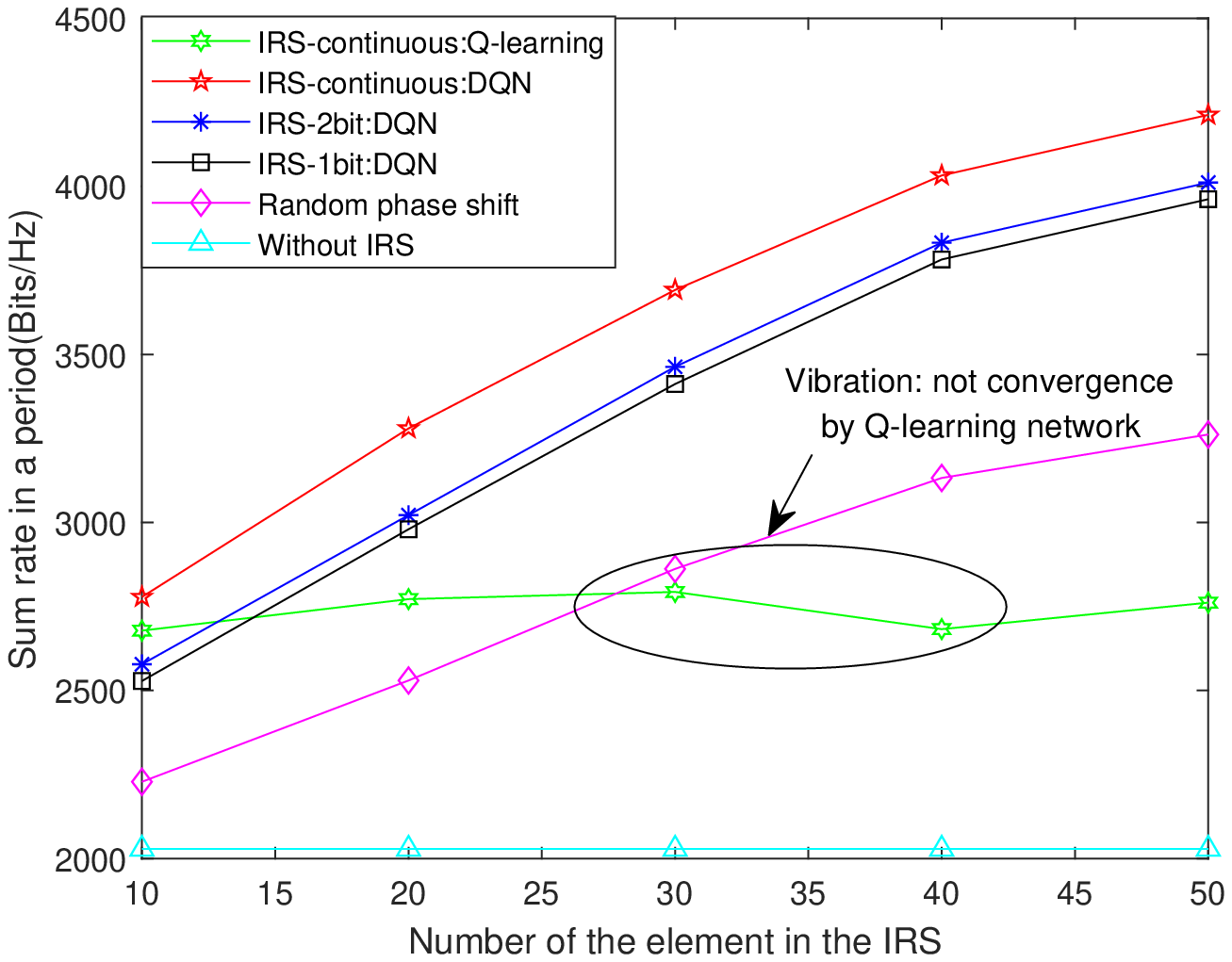}
  \caption{The sum-rate in a period versus Number of IRS Elements $\emph{K}$ in the IRS, $\mathcal{P}$=20dBm, M=5, N=10.}
  \label{Elements}
  \end{minipage}  
\end{figure}

Fig.~\ref{Power} shows the achieved sum-rate versus the transmit power \emph{$\mathcal{P}$} with mentioned different schemes when $\emph{K}$=10, \emph{M}=5, \emph{N}=10. It can be observed that the sum-rate of all schemes increases with the upgrades of the \emph{$\mathcal{P}$}. The performance of the DQN-based optimization algorithm outperforms the Q-learning-based scheme. Furthermore, the sum-rate of the IRS-aided scheme grows faster than that of the ``without IRS" scheme, where the gap between them is expanded to large with the increment of \emph{$\mathcal{P}$}. Compared to the ``Random phase shift" scheme, the performance of DQN-based schemes presents the same growth trend with IRS-aided scheme and without IRS scheme, while the ``Random phase shift" scheme also outperforms the ``without IRS" scheme, these results demonstrate the benefit for the IRS deployment. In the continuous and discrete assumption for IRS phase shift, it can be observed that the discrepancy between them keeps stable with the increase of \emph{$\mathcal{P}$}, and the high-resolution bits guide the small gap to the ideal case, while there is still a big gap for both two discrete cases compared to the ideal case.

\subsubsection{Sum-rate versus Number of IRS Elements}
In Fig.~\ref{Elements}, the achieved sum-rate versus number of elements \emph{$K$} in the IRS with mentioned different schemes when \emph{$\mathcal{P}$}=20dBm, \emph{M}=5, \emph{N}=10. It can be observed that the sum-rate of users increases with the increase of \emph{$K$} under IRS-aided schemes, while the sum-rate for ``without IRS" keeps constant. This is because the large number of elements in the IRS is employed, the higher gain can be achieved in the system. Moreover, the gap grows when the elements \emph{K} increases, while the growth rate for all IRS-aided schemes begins to slow down with the increase the elements \emph{K}. It implies that for serving multiple users, a large number of elements can be considered to apply, however, the high-resolution bits need to be defined in such scenarios. Similarly, for the Q-learning-based scheme, it shows the same vibration situation comparable to the transmit power \emph{$\mathcal{P}$}, which also indicates that DQN-based schemes outperform the Q-learning-based scheme.

\subsection{Impact of decoding order}
\begin{figure}[htbp]
  \centering
  \begin{minipage}[t]{0.48\textwidth}
  \centering
  \includegraphics[height=2.4in,width=3.2in]{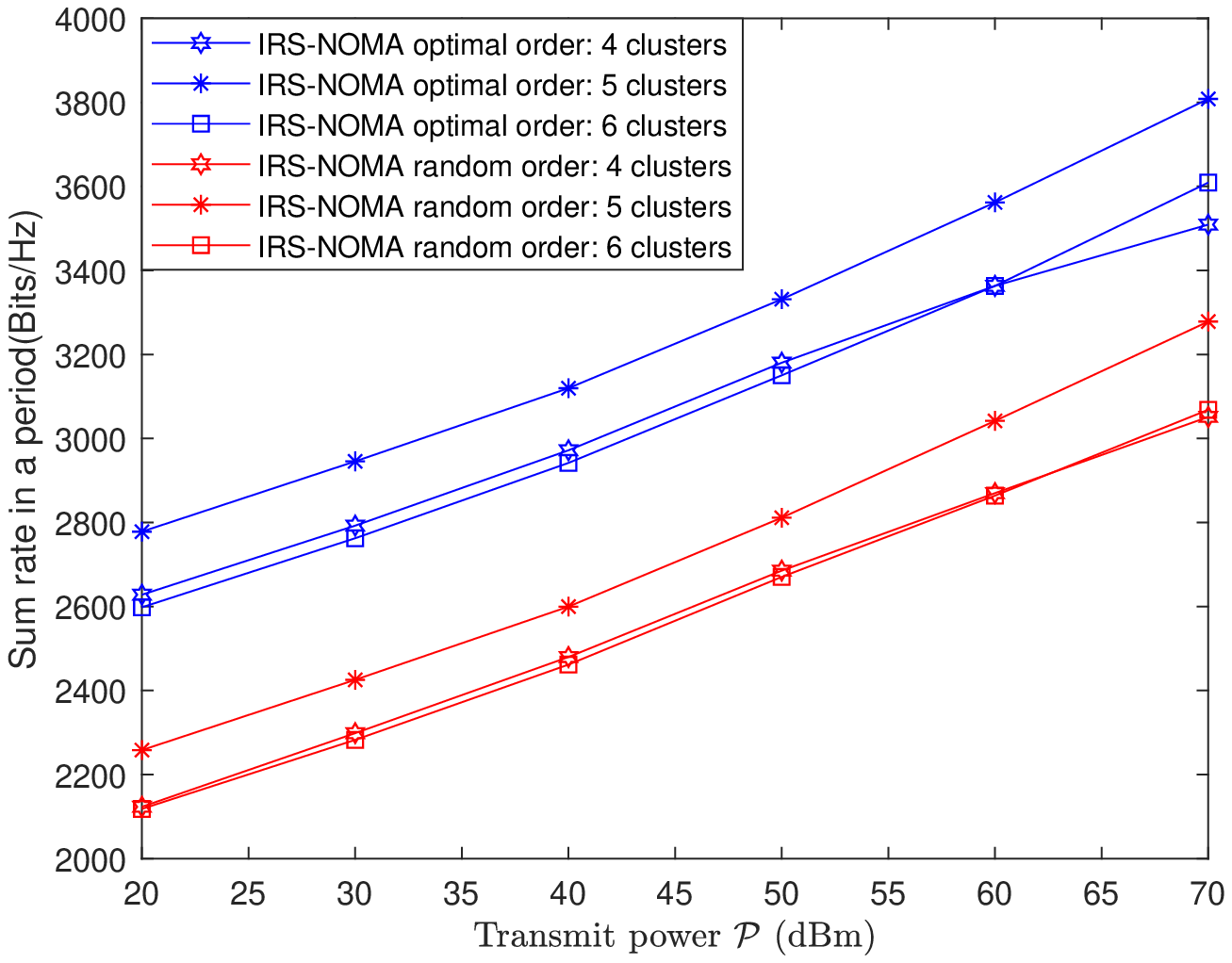}
  \caption{Sum rate in a period with different decording order, K=10, N=10.}
  \label{Order}
  \end{minipage}
  \begin{minipage}[t]{0.48\textwidth}
  \centering
  \includegraphics[height=2.4in,width=3.2in]{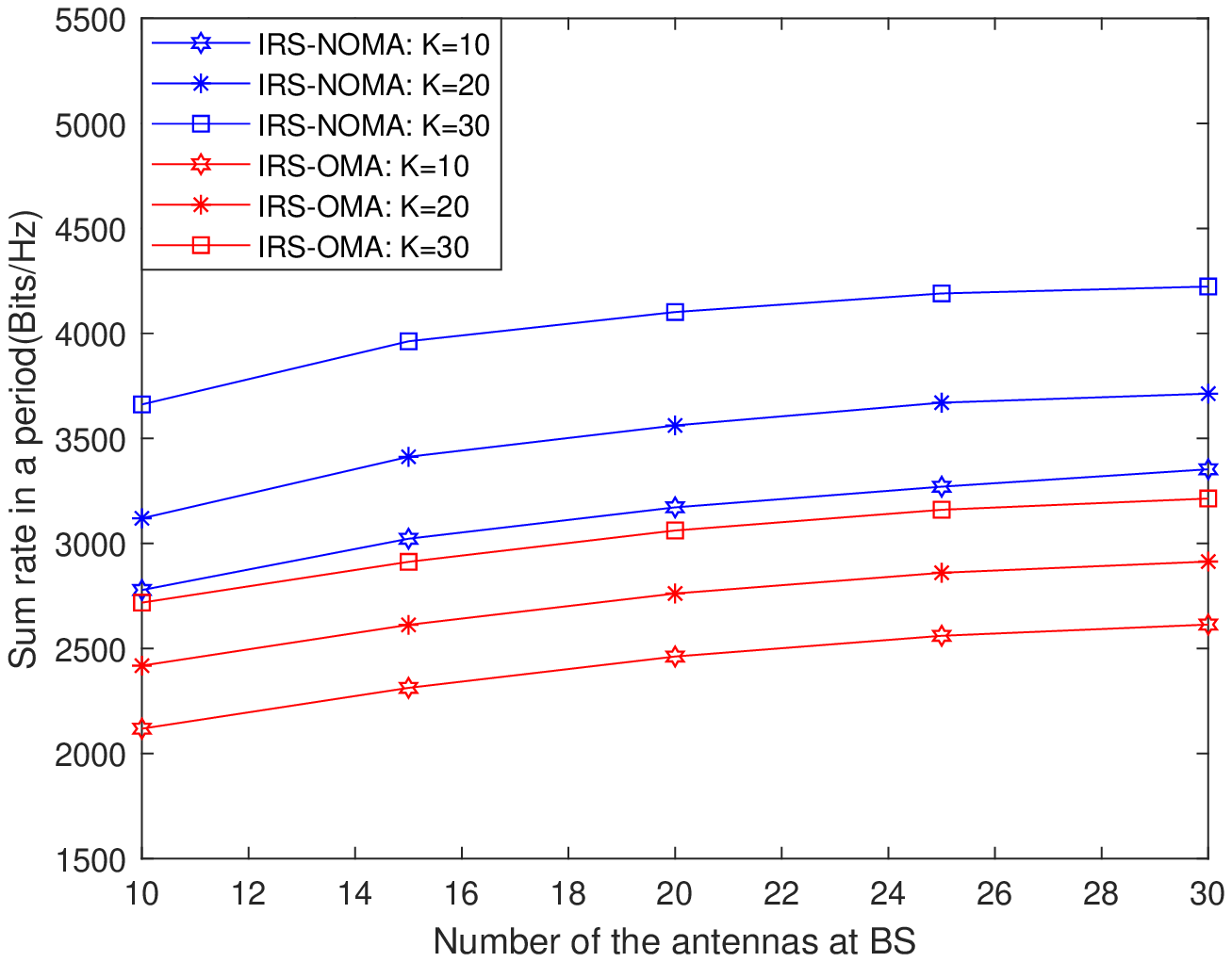}
  \caption{Comparison IRS-NOMA scheme with IRS-OMA scheme.}
  \label{NOMAOMA}
  \end{minipage}  
\end{figure}

In order to explore the influence brought by decoding order, we compare two schemes based on the different numbers of clusters: optimal order and random order. Optimal order denotes the best order selected by the DQN-based algorithm, while the random order represents the stochastic decoding sequence for users. As illustrated in Fig.~\ref{Order}, the DQN-based optimal order scheme significantly outperforms the random order scheme, which highlights finding the optimal decoding order. Additionally, it can be observed that the gaps among the random order schemes present a slighter change compared to the optimal order schemes. This is because optimal order schemes provide the extreme values in each transmit power, which is ignored in random order schemes. Note that the optimal decoding order is found by exhaustive search algorithm, which costs over \emph{L}! iterations, which can only be adopted when the number of users is not large, otherwise, it will bring high complexity for the whole algorithm.

\subsection{Comparison between NOMA and OMA}
Finally, we compare the performance of the IRS-NOMA scheme to that of the IRS-OMA scheme. In the IRS-OMA scheme, the communication system follows the TDMA with the aid of the IRS. Under the case of three different number of elements in the IRS \emph{K}=10, 20, 30, Fig.~\ref{NOMAOMA} characterizes the performance of both schemes. It is observed that when all users are served with the same transmit power, the IRS-NOMA scheme outperforms the IRS-OMA scheme with the increase of the number of the antenna at BS, and the gaps between IRS-NOMA schemes also achieve higher difference than that of IRS-OMA schemes. This is because the NOMA can serve users simultaneously compared to the OMA scheme.

\section{Conclusion}
\par
In this paper, we explored the IRS-aided MISO-NOMA network. The sum-rate maximization problem was formulated by jointly optimizing the passive beamforming vector at the IRS, decoding order, power allocation coefficient vector, and the number of clusters under QoS constraint. To tackle the problem formulated, three machine learning algorithms were proposed to predict user mobility, partition users into clusters and design the phase shift matrix, respectively. Numerical results were provided for demonstrating that the proposed IRS-NOMA scheme achieved significant performance gain compared to the IRS-OMA scheme. Additionally, properly choosing parameters in the proposed algorithms is capable of improving the training performance of neural networks.

\appendices
\section*{Appendix~A: Proof of Proposition \ref{proposition 1}} \label{Appendix:A}
\renewcommand{\theequation}{A.\arabic{equation}}
\setcounter{equation}{0}

\par
Suppose that there are two users \emph{$l_{i}$} and \emph{$l_{j}$} in the \emph{m}-th cluster with optimal decoding order \emph{$\Omega_{m,l_{j}} > \Omega_{m,l_{i}}$}, and the equation \eqref{9} simplified as \eqref{A}, 
\begin{figure*}
\begin{align}\label{A}
  \frac{\mid\pmb{\upsilon}^{H}\pmb{\Phi}_{m,l_{i}}\pmb{\omega}_{m}\alpha_{m,\tilde{l}_{i}}\mid^2}{\sum\limits_{\Omega_{m,l_{\overline{j}'}}>\Omega_{m,\tilde{l}_{i}}}\mid\pmb{\upsilon}^{H}\pmb{\Phi}_{m,l_{i}}\pmb{\omega}_{m}\alpha_{m,l_{\overline{j}'}}\mid^2+\delta^{2}} \geq \frac{\mid\pmb{\upsilon}^{H}\pmb{\Phi}_{m,\tilde{l}_{i}}\pmb{\omega}_{m}\alpha_{m,\tilde{l}_{i}}\mid^2}{\sum\limits_{\Omega_{m,l_{\overline{j}'}}>\Omega_{m,\tilde{l}_{i}}}\mid\pmb{\upsilon}^{H}\pmb{\Phi}_{m,\tilde{l}_{i}}\pmb{\omega}_{m}\alpha_{m,l_{\overline{j}'}}\mid^2+\delta^{2}}, \nonumber
\end{align}
\begin{align}
  \Rightarrow \hspace{1em} \mid\pmb{\upsilon}^{H}\pmb{\Phi}_{m,l_{i}}\pmb{\omega}_{m}\alpha_{m,\tilde{l}_{i}}\mid^2 &\bigg [\sum\limits_{\Omega_{m,l_{\overline{j}'}}>\Omega_{m,\tilde{l}_{i}}}\mid\pmb{\upsilon}^{H}\pmb{\Phi}_{m,\tilde{l}_{i}}\pmb{\omega}_{m}\alpha_{m,l_{\overline{j}'}}\mid^2+\delta^{2} \bigg ] \nonumber \\
  &\geq \mid\pmb{\upsilon}^{H}\pmb{\Phi}_{m,\tilde{l}_{i}}\pmb{\omega}_{m}\alpha_{m,\tilde{l}_{i}}\mid^2 \bigg [\sum\limits_{\Omega_{m,l_{\overline{j}'}}>\Omega_{m,\tilde{l}_{i}}}\mid\pmb{\upsilon}^{H}\pmb{\Phi}_{m,l_{i}}\pmb{\omega}_{m}\alpha_{m,l_{\overline{j}'}}\mid^2+\delta^{2} \bigg ], \nonumber
\end{align}
\begin{align}
  \Rightarrow \hspace{1em} &\mid\pmb{\upsilon}^{H}\pmb{\Phi}_{m,l_{i}}\pmb{\omega}_{m}\alpha_{m,\tilde{l}_{i}}\mid^2 \bigg [\mid\pmb{\upsilon}^{H}\pmb{\Phi}_{m,\tilde{l}_{i}}\pmb{\omega}_{m}\alpha_{m,\tilde{l}_{i}}\mid^2 I_{l_{i}} + \sum\limits_{\Omega_{m,l_{\overline{j}'}}>\Omega_{m,\tilde{l}_{i}}}\mid\pmb{\upsilon}^{H}\pmb{\Phi}_{m,\tilde{l}_{i}}\pmb{\omega}_{m}\alpha_{m,l_{\overline{j}'}}\mid^2+\delta^{2} \bigg ] \nonumber \\
  &\geq \mid\pmb{\upsilon}^{H}\pmb{\Phi}_{m,\tilde{l}_{i}}\pmb{\omega}_{m}\alpha_{m,\tilde{l}_{i}}\mid^2 \bigg [\mid\pmb{\upsilon}^{H}\pmb{\Phi}_{m,l_{i}}\pmb{\omega}_{m}\alpha_{m,\tilde{l}_{i}}\mid^2 I_{l_{i}} + \sum\limits_{\Omega_{m,l_{\overline{j}'}}>\Omega_{m,\tilde{l}_{i}}}\mid\pmb{\upsilon}^{H}\pmb{\Phi}_{m,l_{i}}\pmb{\omega}_{m}\alpha_{m,l_{\overline{j}'}}\mid^2+\delta^{2} \bigg ],
\end{align}
\hrulefill \vspace*{0pt}
\end{figure*}
where \emph{$I_{l_{i}}$} denotes the intra-interference for user \emph{$l_{i}$} according to the decoding orders, it can be expressed as
\begin{align}
  I_{l_{i}} = \sum_{\Omega_{m,l_{\overline{j}''}}>\Omega_{m,l_{i}}}\mid\pmb{\upsilon}^{H}\pmb{\Phi}_{m,l_{i}}\pmb{\omega}_{m}\alpha_{m,l_{\overline{j}''}}\mid^2
\end{align}
\par
Thus, by rearranging the inequality, it can be obtained that
\begin{align}
R_{m,l_{j} \rightarrow m,l_{i}} \geq R_{m,l_{i} \rightarrow m,l_{i}} \geq R_{m,\tilde{l}_{i}}.
\end{align}
\par
Thus, we can get that the $R_{m,l_{j} \rightarrow m,l_{i}} \geq R_{m,l_{i} \rightarrow m,l_{i}}$ is the necessary condition of $R_{m,l_{i} \rightarrow m,l_{i}} \geq R_{m,\tilde{l}_{i}}$.

\section*{Appendix~C: Proof of Lemma 1} \label{Appendix:C}
\renewcommand{\theequation}{C.\arabic{equation}}
\setcounter{equation}{0}

\par
In order to generate the initial positions, the range of movement for users needs to be determined. In the proposed system, the position of IRS is acted as the origin point to establish a two-dimensional Cartesian coordinate system, where \emph{$\mathcal{M}(x)$} denotes the X-Y plane. In order to make the problem simple, let \emph{$\mathcal{\pi}(x)$} denote the movement range of all users. According to the acceptance-rejection sampling method, the indicator \emph{$\mathcal{I}_{x_{0},y_{0}}$} is proposed to evaluate whether it can be accepted or rejected, which can be expressed as
\begin{align}
  \mathcal{I}_{x_{0},y_{0}} = \begin{cases} 1,&\text{${\rm if\ (x_{0},y_{0})\ is\ accepted},$} \\ 0, &\text{${\rm if\ (x_{0},y_{0})\ is\ rejected},$} \end{cases} 
\end{align}
and the probability density function for \emph{$x_{0}$} and \emph{$y_{0}$} can be expressed as
\begin{align}
  \overline{p}_{x_{0}}(x) = \mathcal{M}(x),
\end{align}
\begin{align}
  \overline{p}_{y_{0}}(y) & = \overline{P}'_{Y_{0}}(y) = \overline{P}'(Y_{0} \leq y) = \overline{P}'(Z\mathcal{M}(x)\mathcal{U} \leq y) \nonumber \\
  & = \overline{P}'(\mathcal{U} \leq \frac{y}{Z\mathcal{M}(x)}) = (\int_{0}^{\frac{y}{Z\mathcal{M}(x)}}1d\mathcal{U})' \nonumber \\
  & = (\frac{y_{0}}{Z\mathcal{M}(x)})' = \frac{1}{Z\mathcal{M}(x)},
\end{align}
where \emph{$\mathcal{U} ~ (0,1)$} and \emph{Z} is the constant value. Thus, the probability density function for (\emph{$x_{0},y_{0}$}) can be calculated as
\begin{align}
  \overline{p}_{x_{0},y_{0}}(\mathcal{I}_{x_{0},y_{0}}=1) = \frac{\overline{p}_{x_{0}}(x)\overline{p}_{y_{0}}(y)}{\overline{p}(\mathcal{I}_{x_{0},y_{0}}=1)} = \frac{1}{Z}.
\end{align}
\par
Then, the initial positions of the users can be randomly generated according to this probability density function.

\ifCLASSOPTIONcaptionsoff
  \newpage
\fi

\begin{IEEEbiography}[{\includegraphics[width=1in, height=1.25in,clip, keepaspectratio]{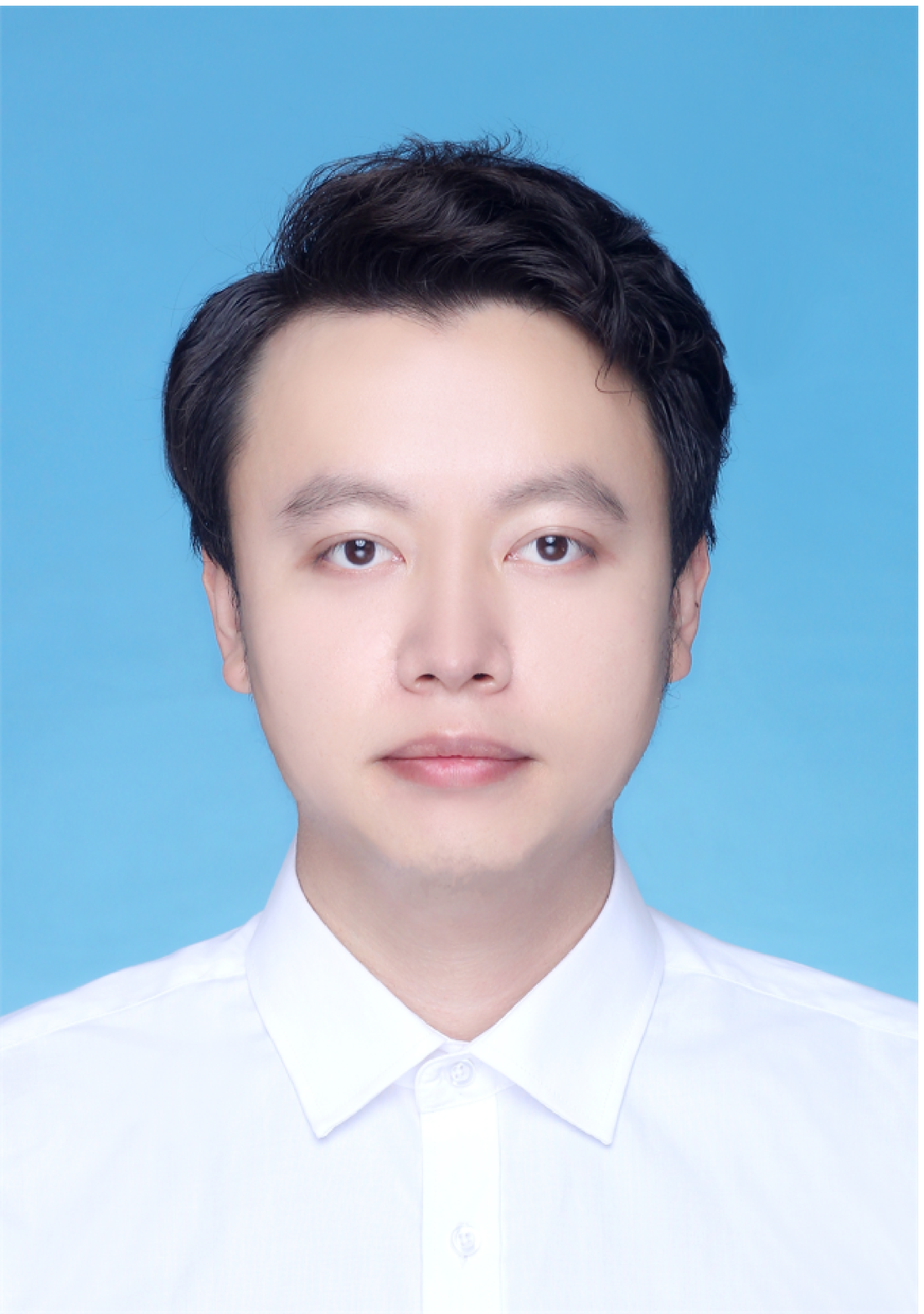}}]
{Xinyu Gao} (Student Member, IEEE) received the B.S. degree and M.S. degree from China Agricultural Univerisity, Beijing, China, in 2016 and 2019, respectively. He is currently working toward the Ph.D. degree with Communication Systems Research Group, School of Electronic Engineering and Computer Science, Queen Mary University of London, U.K.. 
\par
His research interests include machine learning, NOMA techniques, RIS, as well as robot localization and navigation.
\end{IEEEbiography}

\begin{IEEEbiography}[{\includegraphics[width=1in, height=1.25in,clip, keepaspectratio]{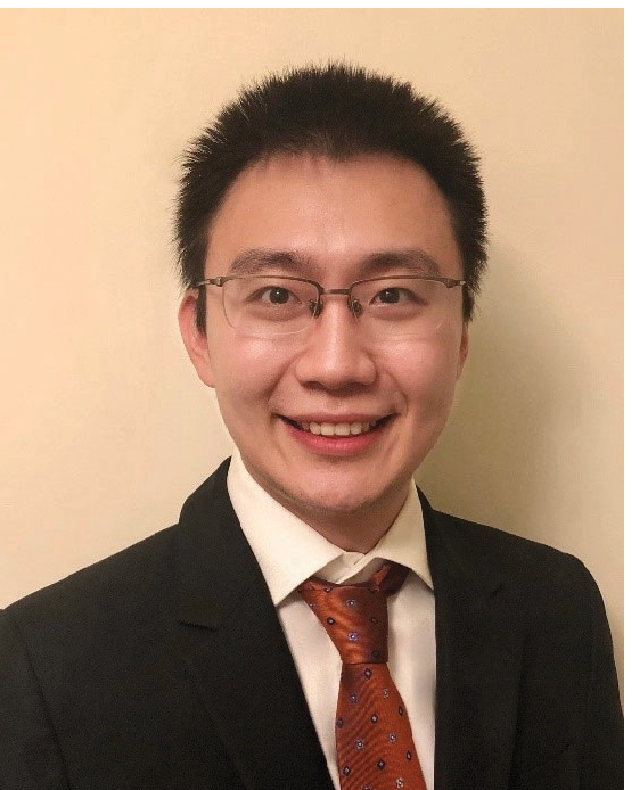}}]
{Yuanwei Liu} (Senior Member, IEEE, \url{http://www.eecs.qmul.ac.uk/~yuanwei}) received the B.S. and M.S. degrees from the Beijing University of Posts and Telecommunications in 2011 and 2014, respectively, and the Ph.D. degree in electrical engineering from the Queen Mary University of London, U.K., in 2016. He was with the Department of Informatics, King's College London, from 2016 to 2017, where he was a Post-Doctoral Research Fellow. He has been a Lecturer (Assistant Professor) with the School of Electronic Engineering and Computer Science, Queen Mary University of London, since 2017. His research interests include non-orthogonal multiple access, 5G/6G networks, machine learning, and stochastic geometry.
\par
Dr. Liu is currently an Editor on the Editorial Board of the \textsc{IEEE Transactions on Wireless Communications}, the \textsc{IEEE Transactions on Communications}, and \textsc{IEEE Communications Letters}. He serves as the leading Guest Editor for \textsc{IEEE JSAC} special issue on Next Generation Multiple Access, a Guest Editor for \textsc{IEEE JSTSP} special issue on Signal Processing Advances for Non-Orthogonal Multiple Access in Next Generation Wireless Networks. He has served as a TPC Member for many IEEE conferences, such as GLOBECOM and ICC. He received IEEE ComSoc Outstanding Young Researcher Award for EMEA in 2020. He received the Exemplary Reviewer Certificate of \textsc{IEEE Wireless Communications Letters} in 2015, \textsc{IEEE Transactions on Communications} in 2016 and 2017, and \textsc{IEEE Transactions on Wireless Communications} in 2017 and 2018. He has served as the Publicity Co-Chair for VTC 2019-Fall. He is the leading contributor for ``Best Readings for Non-Orthogonal Multiple Access (NOMA)'' and the primary contributor for ``Best Readings for Reconfigurable Intelligent Surfaces (RIS)''. He serves as the chair of Special Interest Group (SIG) in Signal Processing and Computing for Communications (SPCC) Technical Committee on the topic of signal processing Techniques for next generation multiple access (NGMA), the vicechair of SIG Wireless Communications Technical Committee (WTC) on the topic of Reconfigurable Intelligent Surfaces for Smart Radio Environments (RISE), and the Tutorials and Invited Presentations Officer for Reconfigurable Intelligent Surfaces Emerging Technology Initiative.
\end{IEEEbiography}

\begin{IEEEbiography}[{\includegraphics[width=1in, height=1.25in,clip, keepaspectratio]{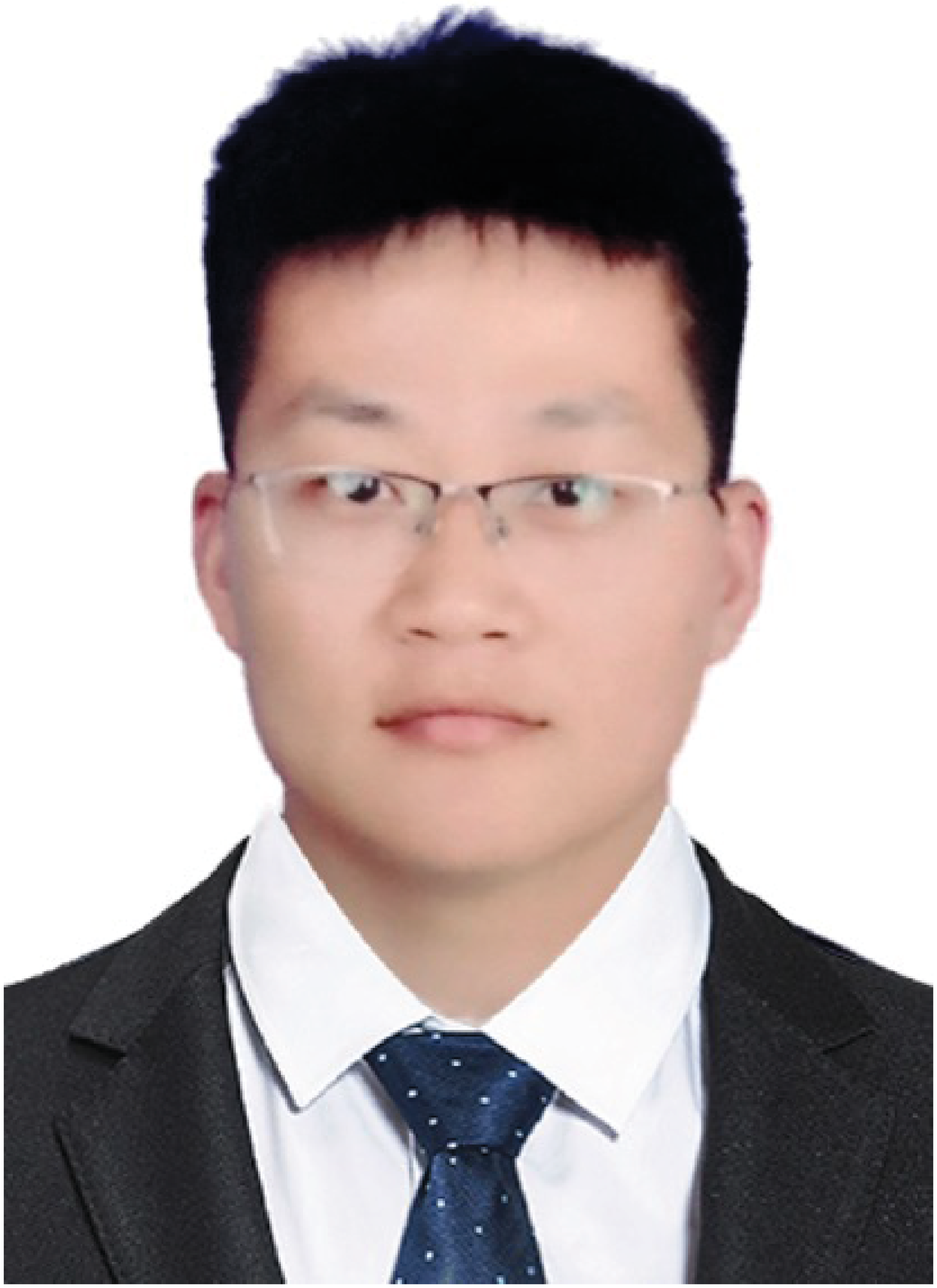}}]
{Xiao Liu} (Student Member, IEEE) is currently working toward the Ph.D. degree with Communication Systems Research Group, School of Electronic Engineering and Computer Science, Queen Mary University of London. His research interests include UAV aided networks, machine learning, NOMA techniques, RIS.
\par
He is the exemplary reviewer of IEEE Transactions on Communication and IEEE Communications Letter, 2019.
\end{IEEEbiography}

\begin{IEEEbiography}[{\includegraphics[width=1in, height=1.25in,clip, keepaspectratio]{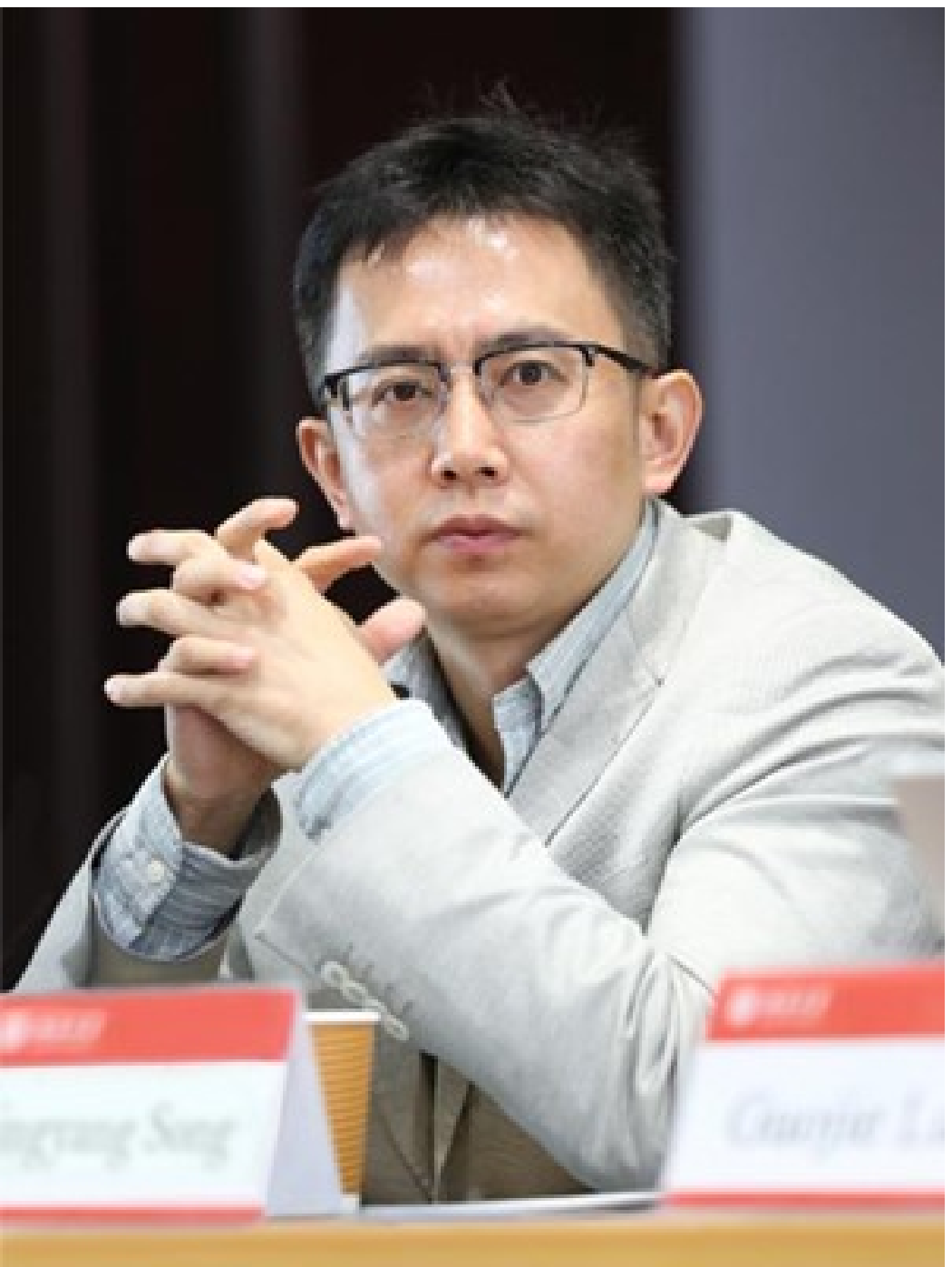}}]
{Lingyang Song} (Fellow, IEEE) received a Ph.D. from the University of York, UK, in 2007, where he received the K. M. Stott Prize for excellent research. He worked as a postdoctoral research fellow at the University of Oslo, Norway, and Harvard University, until rejoining Philips Research UK in March 2008. In May 2009, he joined the School of Electronics Engineering and Computer Science, Peking University, China, as a full professor. His main research interests include MIMO, OFDM, cooperative communications, cognitive radio, physical layer security, game theory, and wireless ad hoc/sensor networks.
\par
He is co-inventor of a number of patents (standard contributions), and author or co-author of over 100 journal and conference papers. He received the best paper award at the IEEE International Conference on Wireless Communications, Networking and Mobile Computing (WiCOM 2007), the best paper award at the 1st IEEE International Conference on Communications in China (ICCC 2012), the best student paper award at the 7th International Conference on Communications and Networking in China (ChinaCom2012), the best paper award at the IEEE Wireless Communication and Networking Conference (WCNC2012), and the best paper award at the International Conference on Wireless Communications and Signal Processing (WCSP 2012).
\par
He is currently on the Editorial Board of IET Communications and the Journal of Network and Computer Applications, and a guest editor ofElsevier Computer Communications and the EURASIP Journal on Wireless Communications and Networking. He serves as a member of the Technical Program Committee and Co-chair for several international conferences and workshops.
\end{IEEEbiography}

\end{document}